\documentclass[10pt]{article}

\usepackage[dvips]{graphicx}
\usepackage{wrapfig}
\usepackage{amssymb}

\def\IZ{\mathbb {Z}}

\def\IR{\mathbb {R}}
\def\IC{\mathbb {C}}
\newcommand{\IP}{{\relax{\rm I\kern-.18em P}}}
\newcommand{\IF}{{\relax{\rm I\kern-.18em F}}}
\newcommand{\II}{I\hspace{-.1em}I}

\renewcommand{\thefootnote}{\fnsymbol{footnote}}

\newdimen\tableauside\tableauside=1.0ex
\newdimen\tableaurule\tableaurule=0.4pt
\newdimen\tableaustep
\def\phantomhrule#1{\hbox{\vbox to0pt{\hrule height\tableaurule width#1\vss}}}
\def\phantomvrule#1{\vbox{\hbox to0pt{\vrule width\tableaurule height#1\hss}}}
\def\sqr{\vbox{%
  \phantomhrule\tableaustep
  \hbox{\phantomvrule\tableaustep\kern\tableaustep\phantomvrule\tableaustep}%
  \hbox{\vbox{\phantomhrule\tableauside}\kern-\tableaurule}}}
\def\squares#1{\hbox{\count0=#1\noindent\loop\sqr
  \advance\count0 by-1 \ifnum\count0>0\repeat}}
\def\tableau#1{\vcenter{\offinterlineskip
  \tableaustep=\tableauside\advance\tableaustep by-\tableaurule
  \kern\normallineskip\hbox
    {\kern\normallineskip\vbox
      {\gettableau#1 0 }%
     \kern\normallineskip\kern\tableaurule}%
  \kern\normallineskip\kern\tableaurule}}
\def\gettableau#1 {\ifnum#1=0\let\next=\null\else
  \squares{#1}\let\next=\gettableau\fi\next}

\makeatletter
 \renewcommand{\theequation}{%
       \thesection.\arabic{equation}}
 \@addtoreset{equation}{section}
\makeatother

\newcommand{\qed}{\hbox{\rule[-2pt]{3pt}{6pt}}}

\makeatletter
\def\eqnarray{%
 \stepcounter{equation}%
 \let\@currentlabel=\theequation
 \global\@eqnswtrue
 \global\@eqcnt\z@
 \tabskip\@centering
 \let\\=\@eqncr
 $$\halign to \displaywidth\bgroup\@eqnsel\hskip\@centering
 $\displaystyle\tabskip\z@{##}$&\global\@eqcnt\@ne
 \hfil$\displaystyle{{}##{}}$\hfil
 &\global\@eqcnt\tw@$\displaystyle\tabskip\z@{##}$\hfil
 \tabskip\@centering&\llap{##}\tabskip\z@\cr}
\makeatother

\begin{document}
\begin{titlepage}
 \begin{center}
\vspace*{8em}{\bf\LARGE Topological open string amplitudes on local toric}\\[1em]
{\bf\LARGE del Pezzo surfaces via remodeling the B-model}\\[3.5em]
{\Large Masahide Manabe\footnote[1]{e-mail: d07002p@math.nagoya-u.ac.jp}}\\[3em]
{\em\Large Graduate School of Mathematics, Nagoya University,}\\[0.5em]
{\em\Large Nagoya, 464-8602, Japan}\\[9em]
{\bf\large Abstract}\\[1em]
 \end{center}
We study topological strings on local toric del Pezzo surfaces by a method called remodeling the B-model which was recently proposed by Bouchard, Klemm, Mari\~no and Pasquetti. For a large class of local toric del Pezzo surfaces we prove a functional formula of the Bergman kernel which is the basic constituent of the topological string amplitudes by the topological recursion relation of Eynard and Orantin. Because this formula is written as a functional of the period, we can obtain the topological string amplitudes at any point of the moduli space by a simple change of variables of the Picard-Fuchs equations for the period. By this formula and mirror symmetry we compute the A-model amplitudes on $K_{{\IF}_2}$, and predict the open orbifold Gromov-Witten invariants of ${\IC}^3/{\IZ}_4$.
\end{titlepage}
%\tableofcontents
\renewcommand{\thefootnote}{\arabic{footnote}}
\section{Introduction}%sec1

From a viewpoint of string compactification \cite{CHSW}, the research of Calabi-Yau threefold has been made by mathematicians and especially physicists since twenty-five years ago. Let us consider topological string theory on a Calabi-Yau threefold, then from the topological string amplitudes we can obtain many informations such as a part of $F$-term in the low energy effective action of four dimensional ${\cal N}=2$ supergravity \cite{BCOV1,AGNT1,AGNT2}, and the number of BPS bound states of D0 and D2 branes which is called the Gopakumar-Vafa invariant \cite{GoVa1,GoVa2} and so on. On the other hand, by geometric engineering \cite{KKVa}, we can embed four (or five) dimensional gauge theories with eight supercharges into type {\II}A superstring (or M) theory, especially by using a family of local toric Calabi-Yau threefolds giving rise to these gauge theories, we can obtain informations about the four (or five) dimensional $SU(N)$ gauge theories with eight supercharges. In \cite{IqKP1,IqKP2,EgKn1,EgKn2} it was proved that an exact agreement between the partition functions of the topological A-model on the above family of local toric Calabi-Yau threefolds and the Nekrasov's formula \cite{Nek} for the supersymmetric $SU(N)$ gauge theories on ${\IR}^4\times S^1$. When we consider the topological A-model on a local toric Calabi-Yau threefold, we find the stringy region where $\alpha '$-correction becomes important and we see that there are orbifold phases as the stringy region. Mathematically this is studied in the context of what is called the crepant resolution conjecture \cite{Coat}. Furthermore we can also consider the case that there are insertions of A-branes to edges of the toric (web) diagram. In this paper we concentrate on computing the A-model amplitudes on these background geometries. Basically two computation are well known ;

$\bullet$ Direct calculus by making use of the topological vertex (or the localization theorem on the torus fixed points) is easily carried out for the large radius phase of local toric Calabi-Yau threefold \cite{AgKMVa}.

$\bullet$ By the mirror symmetry and the BCOV holomorphic anomaly equation \cite{BCOV1}, we can compute the topological closed string amplitudes not only for the large radius phase but also for the orbifold phase \cite{COGP}. Recently Walcher proposed the holomorphic anomaly equation with frozen open string moduli \cite{Wal2}, but at present we cannot compute the topological string amplitudes with genuine open string moduli in this method.

\begin{figure}[htbp]
 \begin{center}
  \includegraphics[width=140mm]{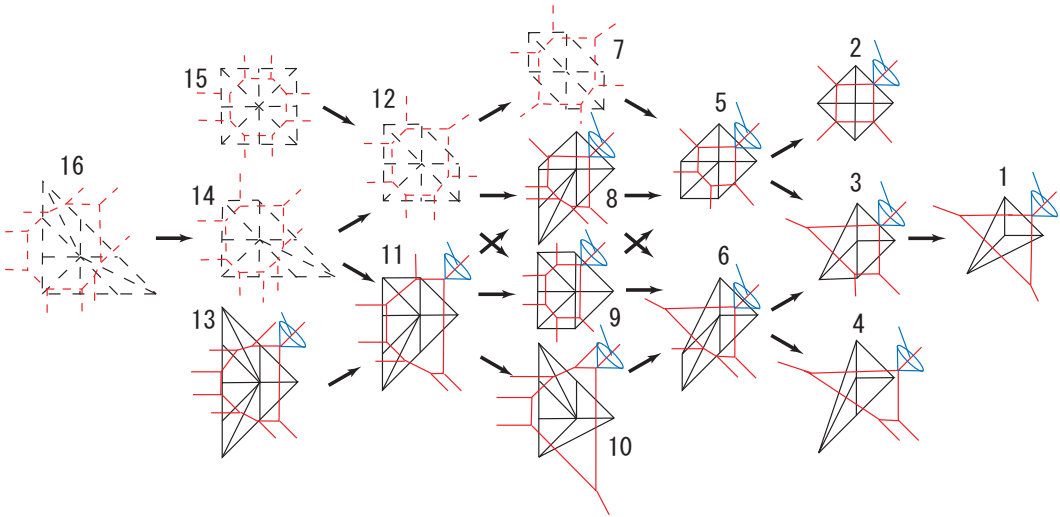}
 \end{center}
 \caption{Batyrev's classification of two dimensional reflexive polytopes and their blow-downs \cite{Bat,Mina}: In this paper we consider $11$ reflexive polytopes described by the solid lines. Here we denoted the fan by the black line, the $(p,q)$-web diagram by the red line and an inserted A-brane considered in section 4 by the blue line. We have numbered the diagrams according to \cite{CKYZ}.}
 \label{fig:1}
\end{figure}

Recently Bouchard, Klemm, Mari\~no and Pasquetti conjectured \cite{Mar1,BKMP1} that for local toric Calabi-Yau threefold, the A-model amplitudes with genuine open string moduli of A-branes inserted to edges of the toric (web) diagram are obtained by mirror symmetry and the topological recursion relation recently proposed by Eynard and Orantin \cite{EO1}. In \cite{EMO}, by discussing modularity it was proved that non-holomorphic amplitudes obtained from this recursion relation satisfy the BCOV holomorphic anomaly equation, and in \cite{DV2} this topological recursion relation was rederived from the viewpoint of two dimensional Kodaira-Spencer theory. In this paper we study the topological A-model on 11 local toric del Pezzo surfaces described by the solid lines in figure \ref{fig:1}. Our main result is that the Bergman kernel (annulus amplitude) obtained from (\ref{eqn:4.6}) in section 4 (and (\ref{eqn:2.13})) as an extension of the formula obtained in the case of local ${\IP}^2$ \cite{BKMP2}. Because this formula is written as a functional of the period which is a solution to the Picard-Fuchs equations, we can obtain the annulus amplitude at any point of the moduli space by expanding the solutions to the Picard-Fuchs equations in an appropriate coordinate. Furthermore since the Bergman kernel is the basic constituent of the Eynard-Orantin's topological recursion relation, we can also compute the higher amplitudes at any point of the moduli space. We hope that this result sheds light on the BKMP conjecture (\ref{eqn:3.13}) and the structure of the holomorphic anomaly equation with open string moduli.

This paper is organized as follows. In section 2 we summarize the recursion relation proposed by Eynard and Orantin and we write this in the form proposed in \cite{BKMP2}. In section 3 we review the topological string amplitudes that we consider in this paper. Section 4 is the main part of this paper, we prove (\ref{eqn:4.6}) as a functional formula of the annulus amplitudes on the above 11 local toric del Pezzo surfaces. We apply this formula for several examples, and especially predict the open orbifold Gromov-Witten invariants of ${\IC}^3/{\IZ}_4$. We discuss the one-holed torus amplitude and the genus zero, three-hole amplitude in section 5. Section 6 is the conclusion of this paper. In appendix A we describe a calculus of the A-model amplitudes by the topological vertex \cite{AgKMVa}. In appendix B we study the torus amplitudes in several examples and check the consistency with the BCOV holomorphic anomaly equation \cite{BCOV2,KlZas}. In appendix C we summarize the transformations of open string moduli proposed in \cite{BKMP1}, and consider the framing ambiguity of an inserted A-brane. In appendix D we summarize the one-holed torus amplitudes on $K_{{\IF}_0}$ and $K_{{\IF}_2}$ on the mirror side. In appendix E we summarize the topological open string amplitudes on $K_{{\IF}_2}$ and ${\IC}^3/{\IZ}_4$.

\section{Eynard-Orantin's topological recursion relation}%sec2

In this section we summarize the topological recursion relation proposed by Eynard and Orantin \cite{EO1}. Let us consider a genus ${\bar g}$ Riemann surface with punctures
\begin{equation}
\Sigma_{\bar{g}}=\left\{x,y \in {\IC}^*~|~H(x,y;z_{\alpha})=0\right\}~\subset {\IC}^* \times {\IC}^*~,
\label{eqn:2.1}%ラベル指定
\end{equation}
where $z_{\alpha},~{\alpha}=1,\cdots,n$ are the deformation parameters of the complex structure of $\Sigma_{\bar g}$. We call the curve $H(x,y;z_{\alpha})=0$ the spectral curve. By $q_i, i=1,\cdots,2{\bar g}+2$ let us denote the ramification points of $H(x,y;z_{\alpha})=0$. On neighborhood of $q_i$, we find two distinct points $q,{\bar q}\in \Sigma_{\bar g }$ such that $x(q)=x(\bar{q})$ by a projected coordinate. On $\Sigma_{\bar{g}}$, Bouchard, Klemm, Mari\~no and Pasquetti defined the free energies $F^{(g,h)}(p_1,\cdots,p_h)$, $g,h \in {\IZ}_{\ge 0},~h \ge 1,~p_i\in \Sigma_{\bar{g}}$ as follows \cite{BKMP1} ;
\begin{eqnarray}
\label{eqn:2.2}%ラベル指定
&&F^{(g,h)}(p_1,\cdots,p_h):=\int^{p_1}\cdots \int^{p_h} W^{(g,h)}(p_1,\cdots,p_h)~,\\
&&W^{(0,1)}(p):=\omega(p):=\log y(p)\frac{dx(p)}{x(p)}~,\quad W^{(0,2)}(p_1,p_2):=B(p_1,p_2)-\frac{dp_1dp_2}{(p_1-p_2)^2}~, \nonumber\\
&&W^{(g,h)}(p_1,\cdots,p_h):=\widetilde{W}^{(g,h)}(p_1,\cdots,p_h)~,\quad (g,h)\neq (0,1),(0,2)~, \nonumber
\end{eqnarray}
where the above integral is carried out by a chain integral $[p_i,p^*_i]$ for each points $p_i$ and certain fixed points $p^*_i$. $B(p_1,p_2)$ is the Bergman kernel defined in the following. $\widetilde{W}^{(g,h)}(p_1,\cdots,p_h)$ is a multilinear meromorphic differential defined by the Eynard-Orantin's topological recursion relation \cite{EO1} inspired from the loop equation of the Hermitian one-matrix model \cite{ACKM,Ake,Eyn1}.
\begin{eqnarray}
&&\widetilde{W}^{(0,1)}(p):=0~,\quad \widetilde{W}^{(0,2)}(p,q):=B(p,q)~, \nonumber\\
&&\widetilde{W}^{(g,h+1)}(p,p_1,\cdots,p_h):=\sum_{q_i}\mathop{Res}_{q=q_i} \frac{dE_{q,\bar{q}}(p)}{\omega(q)-\omega(\bar{q})}\Biggl\{\widetilde{W}^{(g-1,h+2)}(q,\bar{q},p_1,\cdots,p_h) \nonumber\\
&&\hspace{15em} +\sum_{l=0}^g \sum_{J \subset H}\widetilde{W}^{(g-l,|J|+1)}(q,p_J)\widetilde{W}^{(l,|H|-|J|+1)}(\bar{q},p_{H \backslash J})\Biggl\}~, \nonumber\\ \label{eqn:2.3}%ラベル指定
&&\quad\quad  H=\{1,\cdots,h\}~,~J=\{i_1,\cdots,i_j\}\subset H~,~p_J=\{p_{i_1},\cdots,p_{i_j}\}~, \\
&&dE_{q,\bar{q}}(p):=\frac12\int_q^{\bar{q}}B(p,\xi)~,~\mbox{near a ramification point}~q_i~. \nonumber
\end{eqnarray}
We define the Bergman kernel by the following conditions.
\begin{eqnarray}
&&\bullet~~B(p,q)\mathop{\sim}_{p \to q} \frac{dpdq}{(p-q)^2}+\mbox{finite}~.\quad\quad \bullet~~\mbox{Holomorpic except}~p=q~. \nonumber\\
&&\bullet~~\oint_{A_I}B(p,q)=0~,\quad I=1,\cdots,{\bar g}~,
\label{eqn:2.4}%ラベル指定
\end{eqnarray}
where $(A_I,B^I)$ is a canonical basis of one-cycles on $\Sigma_{\bar{g}}$ except cycles with trivial monodromy. Now we can write the Bergman kernel explicitly for ${\bar g}=0$ and $1$.
In the case of ${\bar g}=0$ Riemann surface which is described by $x(p)=p,~y(p)=y(x)$ (: a rational function), the Bergman kernel is given by 
\begin{equation}
B(p_1,p_2)=\frac{dy_1dy_2}{(y_1-y_2)^2}~,\quad y_i=y(x_i)~.
\label{eqn:2.5}%ラベル指定
\end{equation}
Next, we consider the ${\bar g}=1$ Riemann surface $\Sigma$ which is described by 
\begin{eqnarray}
\label{eqn:2.6}%ラベル指定
&& x(p)=x({\bar p})~,\quad y(p),y({\bar p})=\frac{1}{2}(f(x)\pm b(z_{\alpha})\sqrt{\sigma(x)})~, \\
&& \sigma(x)=\prod_{i=1}^4(x-s_i)=x^4-S_1(z_{\alpha})x^3+S_2(z_{\alpha})x^2-S_3(z_{\alpha})x+S_4(z_{\alpha})~,
\label{eqn:2.7}%ラベル指定
\end{eqnarray}
where $b(z_{\alpha})$ is a rational function of $z_{\alpha}$'s and $\Re s_1 \leq \Re s_2 \leq \Re s_3 \leq \Re s_4$, and $f(x)$ is a polynomial of degree two or less than two. In this case the Bergman kernel is given by the Akemann's formula \cite{Ake} which is derived from two-cut ansatz of the Hermitian one-matrix model.
\begin{eqnarray}
\label{eqn:2.8}%ラベル指定
&&B(x_1,x_2)=\frac{dx_1dx_2}{2(x_1-x_2)^2}+\frac{dx_1dx_2}{4\sqrt{\sigma(x_1)\sigma(x_2)}}\left\{\frac{M(x_1,x_2)+M(x_2,x_1)}{(x_1-x_2)^2}-\frac{E(k)}{K(k)}(s_1-s_3)(s_2-s_4)\right\}~,\hspace{3em}\\
\label{eqn:2.9}%ラベル指定
&&\quad\quad M(x_1,x_2)=(x_1-s_1)(x_1-s_2)(x_2-s_3)(x_2-s_4)~,\\
&&\quad\quad K(k)=\int_0^1\frac{dt}{\sqrt{(1-t^2)(1-k^2t^2)}}~,\quad E(k)=\int_0^1dt \sqrt{\frac{1-k^2t^2}{1-t^2}}~,\quad k^2=\frac{(s_1-s_2)(s_3-s_4)}{(s_1-s_3)(s_2-s_4)}~.
\label{eqn:2.10}%ラベル指定
\end{eqnarray}
In \cite{BKMP2}, this formula was rewritten as follows. Firstly we get
\begin{eqnarray}
&&M(x_1,x_2)+M(x_2,x_1)=2x_1^2x_2^2-x_1x_2(x_1+x_2)S_1-(x_1+x_2)S_3+2S_4 \hspace{8em}\nonumber\\
&&\hspace{15em}+(s_1s_2+s_3s_4)(x_1^2+x_2^2)+2(s_1+s_2)(s_3+s_4)x_1x_2~,
\label{eqn:2.11}%ラベル指定
\end{eqnarray}
in terms of the elementary symmetric polynomials $S_k$. Here we note the formula
\begin{eqnarray}
\label{eqn:2.12}%ラベル指定
&&K(k)E(k)=\pi^2\left(\frac{E_2(\tau)}{12}+\omega_1^2e_3\right)~,\quad \tau:~\mbox{modulus of}~\Sigma~,\\
&&\omega_1:=\frac{2i}{\pi}\frac{K(k)}{\sqrt{(s_1-s_3)(s_2-s_4)}}~,\quad e_3:=\frac{1}{12}\left(S_2-3(s_1s_2+s_3s_4)\right)~,\nonumber
\end{eqnarray}
where $E_2(\tau)$ is the second Eisenstein series which is a quasi-modular holomorphic form of weight $2$. From (\ref{eqn:2.11}) and (\ref{eqn:2.12}), we obtain
\begin{eqnarray}
\label{eqn:2.13}%ラベル指定
B(x_1,x_2)&=&\frac{dx_1dx_2}{\sqrt{\sigma(x_1)\sigma(x_2)}}\left(\frac{\sqrt{\sigma(x_1)\sigma(x_2)}+f(x_1,x_2)}{2(x_1-x_2)^2}+\frac{G(\tau)}{4}\right)~,\hspace{13em}\\\label{eqn:2.14}%ラベル指定
f(x_1,x_2):&=&x_1^2x_2^2-\frac12 x_1x_2(x_1+x_2)S_1+\frac16(x_1^2+4x_1x_2+x_2^2)S_2-\frac12(x_1+x_2)S_3+S_4~,\\
G(\tau):&=&\frac{E_2(\tau)}{3\omega_1^2}~.
\label{eqn:2.15}%ラベル指定
\end{eqnarray}
In section 4 by regarding the modulus $\tau$ as a function of the complex moduli parameters $z_{\alpha}$ we will prove (\ref{eqn:4.6}). From this formula and the recursion relation (\ref{eqn:2.3}) we can obtain the free energies $F^{(g,h)}(x_1,\cdots,x_h)$ explicitly in the case when the spectral curve has the form (\ref{eqn:2.6}). Furthermore, in \cite{BKMP2} by estimating (\ref{eqn:2.3}) it was proved that $W^{(g,h)}(x_1,\cdots,x_h)$, $g,h \in {\IZ}_{\ge 0},~h \ge 1$, $(g,h)\neq (0,1),(0,2)$ can be written as
\begin{eqnarray}
\label{eqn:2.16}%ラベル指定
&&W^{(g,h)}(x_1,\cdots,x_h)=\frac{dx_1\cdots dx_h}{\Delta(z_{\alpha})^{2g-2+h}\prod_{i=1}^h \sqrt{\sigma(x_i)}}\sum_{k=0}^{3g-3+2h}G(\tau)^kf_k^{(g,h)}(x_1,\cdots,x_h)~,\\
&&\quad f_k^{(g,h)}(x_1,\cdots,x_h)=\frac{Q_k^{(g,h)}(x_1,\cdots,x_h)}{\prod_{i=1}^h \sigma(x_i)^{3g-2+h}}~,\nonumber\\
&&\quad Q_k^{(g,h)}(x_1,\cdots,x_h):~\mbox{a finite degree polynomial of $x_i$ and $z_{\alpha}$}~, \nonumber
\end{eqnarray}
where $\Delta(z_{\alpha})=\prod_{i<j}(s_i-s_j)^2$ is the discriminant of the curve (\ref{eqn:2.7}). In section 5 we will discuss the higher amplitudes $W^{(1,1)}(x)$ and $W^{(0,3)}(x_1,x_2,x_3)$, and in appendix B the torus ($g=1,h=0$) amplitude will be discussed.

\section{Topological strings on local toric Calabi-Yau threefold}%sec3

Now we review the topological string amplitudes on local toric Calabi-Yau threefold, and clarify what to be computed and how to compute it \cite{Mar2,Mar1}. Firstly, let us consider the topological A-model on a local toric Calabi-Yau threefold $M_A$, and we insert an A-brane to an edge of the toric (web) diagram as figure \ref{fig:1} of section 1. This is considered as a Lagrangian submanifold ${\cal L}$ of $M_A$ with $\dim H_1({\cal L},{\IZ})=1$ \cite{AgVa}. The topological A-model can be described by holomorhic maps $\phi$ from $\Sigma_{g,h}$, which is a compact Riemann surface with genus $g$ and $h$ boundaries, to the target variety $M_A$ ;
\begin{eqnarray}
\phi:\quad \Sigma_{g,h}\quad &\longrightarrow&\quad M_A :~\mbox{holomorphic map}~,\nonumber\\
\partial \Sigma_{g,h}=\bigcup_{i=1}^h{\cal C}_i\quad &\longrightarrow&\quad {\cal L}\subset M_A~,
\label{eqn:3.1}%ラベル指定
\end{eqnarray}
where the boundary $\partial \Sigma_{g,h}$ of $\Sigma_{g,h}$ consists of $h$ connected components ${\cal C}_i$, and they are mapped to the Lagrangian submanifold ${\cal L}$ of $M_A$. These informations are summarized by the A-model amplitude which has two contibutions as follows ;
\begin{eqnarray}
\label{eqn:3.2}%ラベル指定
&&1. \quad \mbox{Bulk}:\hspace{3.25em} \mbox{This contribution is encoded by}~\phi_*[\Sigma_{g,h}]=\beta \in H_2(M_A,{\cal L})~.\\
&&2. \quad \mbox{Boundary}:\quad \mbox{This contribution is encoded by}~\phi_*[{\cal C}_i]=\omega_i\gamma~,\quad \omega_i\in{\IZ}~,
\label{eqn:3.3}%ラベル指定
\end{eqnarray}
where $\gamma$ is a basis of $H_1({\cal L},{\IZ})$. By taking these into consideration, the A-model amplitude or the generating function ${\cal F}(V)$ of the free energies ${\cal F}_A^{(g,h)}$ is written as
\begin{eqnarray}
\label{eqn:3.4}%ラベル指定
{\cal F}_{\vec{\omega},g}(Q)&=&\sum_{\beta}N_{\vec{\omega},g,\beta}Q^{\beta},~~\vec{\omega}=(\omega_1,\cdots,\omega_h)~, \\\label{eqn:3.5}%ラベル指定
{\cal F}(V)&=&\sum_{g=0}^{\infty}\sum_{h=1}^{\infty}\sum_{\omega_i}g_s^{2g-2+h}{\cal F}_{\vec{\omega},g}(Q)\frac{1}{h!}TrV^{\omega_1}\cdots TrV^{\omega_h} \\
&=&\sum_{g=0}^{\infty}\sum_{h=1}^{\infty}g_s^{2g-2+h}{\cal F}_A^{(g,h)}(X_1,\cdots,X_h;Q)~,\quad g_s:~\mbox{the string coupling constant}~,\nonumber
\end{eqnarray}
where $Q$ denotes the parameters of the K\"ahler moduli space ${\cal M}_A$ of $M_A$, and $V$ is a holonomy matrix of the gauge group $U(\infty)$ on the source A-brane. In the second equality of (\ref{eqn:3.5}), by transforming from ${\vec \omega}$ to the winding base ${\vec k}=(k_1,k_2,\cdots)$, where $k_i$ is the number of $i$ included in ${\vec \omega}$, we have made an identification
\begin{equation}
\prod\frac{1}{k_j!}\left(TrV^j\right)^{k_j}\quad \longleftrightarrow \quad  \prod X_{I_1}^{j}\cdots X_{I_{k_j}}^{j}=X_1^{\omega_1}\cdots X_h^{\omega_h}~,
\label{eqn:3.6}%ラベル指定
\end{equation}
where $h=\sum k_j$, $\sum_{i=1}^h \omega_i=\sum jk_j$. We interpret $X_i$ as the open string moduli of this model. Note that in mathematical literature, $N_{\vec{\omega},g,\beta}$ is nothing but the open Gromov-Witten invariants of the Calabi-Yau variety $M_A$ with the A-brane. In this paper we want to consider a computation of ${\cal F}_A^{(g,h)}$ on 11 local toric del Pezzo surfaces in figure \ref{fig:1}, especially around any point on the K\"ahler moduli space of these variety via the mirror symmetry \cite{COGP,CKYZ,CoKa}. So we will summarize the mirror symmetry, the mirror map, and the mirror curve given by Hori and Vafa \cite{HoVa}. To explain the mirror symmetry, we introduce a mirror variety $M_B$ of $M_A$. This mirror pair satisfies the following homeomorphic relations between the cohomology of $M_A$ and $M_B$.
\begin{equation}
H^1\left(M_A,\Omega^1(M_A)\right)\simeq H^1\left(M_B,\Omega^2(M_B)\right)~,\quad H^1\left(M_A,\Omega^2(M_A)\right) \simeq H^1\left(M_B,\Omega^1(M_B)\right)~,
\label{eqn:3.7}%ラベル指定
\end{equation}
where $\Omega^p(M)$ is the sheaf of the germs of the holomorphic $p$-form on $M$. Furthermore the mirror symmetry asserts that there is a correspondence between the K\"ahler moduli space ${\cal M}_A$ of $M_A$ and the complex moduli space ${\cal M}_B$ of $M_B$, and we can confirm a duality between the A-model on $M_A$ and the B-model on $M_B$. The mirror map connects these two models by
\begin{eqnarray}
\label{eqn:3.8}%ラベル指定
Q\in {\cal M}_A,~X_i\quad &\mathop{\longleftrightarrow}^{\mbox{mirror map}}&\quad z\in {\cal M}_B,~x_i \\
{\cal F}_A^{(g,h)}(X_1,\cdots,X_h;Q)\quad &\longleftrightarrow&\quad {\cal F}_B^{(g,h)}(x_1,\cdots,x_h;z) 
\label{eqn:3.9}%ラベル指定
\end{eqnarray}
where $X_i$ and $x_i$ are the open string moduli of the A-model and the B-model respectively, and ${\cal F}_B^{(g,h)}$ is the free energies of the B-model on $M_B$. Hereafter we will consider the case that $M_A$ is a local toric Calabi-Yau threefold defined by
\begin{eqnarray}
M_A:&=& X/U(1)^{n-3},\quad U(1)^{n-3}~:~X_i~\to~ e^{i \sum_{\alpha}\epsilon_{\alpha}l_i^{\alpha}}X_i~,\quad r_{\alpha} \in {\cal M}_A~,\hspace{12em} \nonumber\\
X:&=& \left\{~(X_1,\cdots,X_n)\in {\IC}^n ~\left|~ \sum_{i=1}^n l_i^{\alpha}|X_i|^2=r_{\alpha},~\sum_{i=1}^n l_i^{\alpha}=0,~\alpha=1,\cdots,n-3,~l_i^{\alpha}\in{\IZ}\right.\right\}~,
\label{eqn:3.10}%ラベル指定
\end{eqnarray}
where $l_i^{\alpha}$ are $n-3$ charge (or Mori) vectors, and the second condition of (\ref{eqn:3.10}) is the Calabi-Yau condition \cite{AgVa}. In this case we can construct the mirror variety $M_B$ as
\begin{eqnarray}
M_B:&=&\left\{~(\omega^+,\omega^-,x_1,\cdots,x_n)\in {\IC}^2\times ({\IC}^*)^n~\left|~ \omega^+\omega^-=\sum_{i=1}^nx_i,~\prod_{i=1}^nx_i^{l_i^{\alpha}}=z_{\alpha},~\sum_{i=1}^n l_i^{\alpha}=0~\right.\right\} \nonumber\\
&=& \left\{~(\omega^+,\omega^-,x,y)\in {\IC}^2\times ({\IC}^*)^2~\left|~ \omega^+\omega^-=H(x,y;z_{\alpha})~\right.\right\}~,
\label{eqn:3.11}%ラベル指定
\end{eqnarray}
where in the second equality, we used the Calabi-Yau condition for the charge vectors and normalized one of the parameters of this variety to a constant. On this mirror variety, we can introduce a no-where vanishing holomorphic three-form $\Omega$ which is a fundamental quantity of Calabi-Yau geometries, and furthermore we can reduce this $\Omega$ to an one-form $\omega$ on the mirror curve $\Sigma_{\bar g}$ defined by $H(x,y;z_{\alpha})=0$ via integrating out procedure \cite{FoJi}. Concretely, when we solve $H(x,y;z_{\alpha})=0$ for $y$, the result of this procedure is
\begin{equation}
\Omega=Res_{H(x,y;z_{\alpha})-\omega^+\omega^-}\left[\frac{d\omega^+d\omega^-dxdy}{(H(x,y;z_{\alpha})-\omega^+\omega^-)xy}\right]\quad \Longrightarrow \quad \omega(x)=\log(y)\frac{dx}{x}~.
\label{eqn:3.12}%ラベル指定
\end{equation}
We quote this curve $H(x,y;z_{\alpha})=0$ as the mirror curve $\Sigma_{\bar g}$. Furthermore a moduli space of the A-brane considered in this paper is described by this mirror curve via the mirror symmetry \cite{AgVa,BKMP1}. On this curve, we can compute the free energies $F^{(g,h)}$ defined in section 2. Here Bouchard, Klemm, Mari\~no and Pasquetti conjectured that \cite{BKMP1}
\begin{equation}
{\cal F}_B^{(g,h)}(x_1,\cdots,x_h;z)=F^{(g,h)}(x_1,\cdots,x_h;z)~,
\label{eqn:3.13}%ラベル指定
\end{equation}
as an extention of Dijkgraaf and Vafa's work that connected the topological B-model on some blowup Calabi-Yau geometries and some Hermitian one-matrix models \cite{DV1,Mar3}. Before going on further let us consider the closed mirror map (\ref{eqn:3.8}). By making use of (\ref{eqn:3.12}) the closed mirror maps are given by flat coordinates on the complex moduli space ${\cal M}_B$ as
\begin{eqnarray}
\label{eqn:3.14}%ラベル指定
T_{\alpha}(z)&=&\oint_{A_{\alpha}}\omega(x)~,\quad Q_{\alpha}=e^{T_{\alpha}(z)}~,\\
{\cal F}^{(0,0)}_{A,\alpha}&=&\frac{\partial {\cal F}_A^{(0,0)}}{\partial T_{\alpha}}=\oint_{B^{\alpha}}\omega(x)~,\quad (A_{\alpha},B^{\alpha}):~\mbox{a canonical basis of}~\Sigma_{\bar{g}}~,
\label{eqn:3.15}%ラベル指定
\end{eqnarray}
where ${\cal F}_A^{(0,0)}$ is the genus $0$ prepotential of the A-model. The open mirror map is given by flat open string parameter
\begin{equation}
U=\int_{u}^{u^*}\omega(x)~,\quad x=e^{u},~X=e^{U}~,
\label{eqn:3.16}%ラベル指定
\end{equation}
where $u^*$ is a point that $\log y$ jumps $2\pi i$ through this integral region. The closed mirror maps (\ref{eqn:3.14}) are obtained from logarithmic solutions to the Picard-Fuchs (PF) equations ${\cal D}_{\alpha}f(z)=0$, where the PF operators \{${\cal D}_{\alpha}$\} are defined by
\begin{equation}
{\cal D}_{\alpha}:=\prod_{l_i^{\alpha}>0}\left(\frac{\partial}{\partial x_i}\right)^{l_i^{\alpha}}-\prod_{l_i^{\alpha}<0}\left(\frac{\partial}{\partial x_i}\right)^{-l_i^{\alpha}}~,
\label{eqn:3.17}%ラベル指定
\end{equation}
by the charge vectors $l_i^{\alpha}$. (\ref{eqn:3.15}) are also obtained from double logarithmic solutions to the (extended) PF equations \cite{FoJi}. The open local mirror map (\ref{eqn:3.16}) is obtained from the open string extension of the PF equation \cite{LerM} which is not explained here. We will explicitly write these mirror maps for several concrete examples in the next section.

\section{Annulus amplitudes on local toric del Pezzo surfaces}%sec4

In this section we consider the annulus amplitude by (\ref{eqn:2.13}). In \cite{BKMP2}, $G(\tau)$ was obtained as the functional of the period in the case of local ${\IP}^2$. As an extension of their result we show a functional formula (\ref{eqn:4.6}) for 11 local toric surfaces in figure \ref{fig:1}. Because this formula is written as a functional of the period, we can compute the annulus amplitude at any point of the moduli space by expanding the solutions to the PF equations in an appropriate coordinate.\footnote{If the Bergman kernel is computed, then by the Eynard-Orantin's topological recursion relation (\ref{eqn:2.3}), in principle we can obtain the higher amplitudes at any point of the moduli space.} This formula also has a suggestive form as a propagator or a two point function of a free boson on a mirror curve \cite{DV2}. By this formula we compute the annulus amplitudes for several concrete examples. These computation not only give explicit examples of the BKMP conjecture (\ref{eqn:3.13}), but also lead to new results of the open orbifold Gromov-Witten invariants as subsection 4.4.

\subsection{Set up of local toric surfaces}%sec4.1

\noindent  Let us concentrate on the local toric del Pezzo surfaces represented by $11$ reflexive polytopes in two dimensions (\cite{Bat,CKYZ}, see figure \ref{fig:1} in section 1) which have more than two automorphisms, because the mirror curves of these varieties have the form (\ref{eqn:2.6}).\footnote{Note that the local toric surfaces represented by $9$ reflexive polytopes of these polytopes are generally local toric del Pezzo surfaces because these surfaces obtained as blow-ups of ${\IP}^2$ at less than four torus-fixed points. But the local toric surfaces represented by remaining $2$ reflexive polytopes (no.$11$ and no.$13$ of figure \ref{fig:1}) are special local toric del Pezzo surfaces which complex moduli of these surfaces are fixed.} To compute the annulus amplitudes we give the data of these geometries. These data are easily obtained from the toric data of the reflexive polytopes, and because $11$ surfaces that we consider in this paper are obtained from the (special) fifth del Pezzo surface $dP_5$ (no.$13$ of figure \ref{fig:1}) by blow-downs, we concentrate on $K_{dP_5}$ (local toric $dP_5$). This is constructed from the following six charge vectors $l_i^{\alpha},~\alpha=0,\cdots 5,~i=1,\cdots, 9$ in (\ref{eqn:3.10}),
\begin{eqnarray}
l^{\alpha}&=&(-\alpha,1,\alpha-2,e^{\alpha+1})~,\quad \alpha=0,\cdots,4~,\nonumber\\
l^5&=&(-2,0,1,e^6)~,\quad e^n_i=\delta_{n,i}~,\quad n,i=1,\cdots,6~.
\label{eqn:4.1}%ラベル指定
\end{eqnarray}
The charge vectors of the above $11$ local toric surfaces can be resumed by the following scheme.
\begin{center}
 \begin{tabular}{lclclclclcl}
  & & & & {\bf 8} & & {\bf 5} & & {\bf 2} & & \\
  & & & & $[l^2,l^3,l^4,l^5]$ & $\longrightarrow$ & $[l^2,l^3,l^5]$ & $\longrightarrow$ & $[l^2,l^5]$ & & \\
{\bf 13}  & & {\bf 11} & $\nearrow$ & {\bf 9} & $\searrow \hspace{-1em}\nearrow$ & {\bf 6} & $\searrow$ & {\bf 3} & & {\bf 1} \\
 $[l^{\alpha},l^5]$ & $\longrightarrow$ & $[l^1,l^2,l^3,l^4,l^5]$ & $\longrightarrow$ & $[l^1,l^2,l^3,l^5]$ & $\longrightarrow$ & $[l^3,l^4,l^5]$ & $\longrightarrow$ & $[l^3,l^5]$ & $\longrightarrow$ & $[l^3]$ \\
  & & & $\searrow$ & {\bf 10} & $\nearrow$ & & $\searrow$ & {\bf 4} & & \\
  & & & & $[l^0,l^1,l^2,l^3]$ & & & & $[l^4,l^5]$ & & 
 \end{tabular}
\end{center}
We find that this scheme surely corresponds to the blow-downs of figure \ref{fig:1}, and here we note $l^3=l^1+l^5$ and $l^4=l^0+2l^5$ in the following examples such as $K_{{\IF}_2}$ (example 3). From (\ref{eqn:3.11}) we can construct a mirror curve $\Sigma$ of $K_{dP_5}$. Here by setting $x=-x_1,~y=-x_2,~x_3=\lambda^{-1}$, $\Sigma$ is described by the equation
\begin{equation}
\Sigma~:\quad \lambda y^2-(1-\lambda x+{\tilde z}_5x^2)y+\lambda^{-1}\sum_{\alpha=0}^4 {\tilde z}_{\alpha}(-x)^{\alpha}=0~,
\label{eqn:4.2}%ラベル指定
\end{equation}
where ${\tilde z}_{\alpha}=\lambda^{\alpha}z_{\alpha}$ for $\alpha=0,\cdots,4$ and ${\tilde z}_5=\lambda^2z_5$. Note that this curve is described by $x$ and $y$, whereas $\lambda$ is considered as a scaling freedom of this curve, and by this setting we can describe the moduli space of the A-brane described in no.$13$ of figure \ref{fig:1} \cite{AgVa,BKMP1}. This curve has the same form as (\ref{eqn:2.6}), so we can use the formula (\ref{eqn:2.13}).

\subsection{A functional formula of the annulus amplitudes}%sec4.2

\noindent Before computing the Bergman kernel, let us note the following important remarks. From (\ref{eqn:2.13}), the Bergman kernel is constructed from $f(x_1,x_2)$ and $G(\tau)$, and these quantities correspond to
\begin{eqnarray}
\label{eqn:4.3}%ラベル指定
&&\mbox{\bf 1.}\quad f(x_1,x_2)~\longleftarrow~\mbox{a holomorphic ambiguity}~,\\
&&\mbox{\bf 2.}\quad G(\tau)~\longleftarrow~\mbox{a bulk (modular) dependent term}~. 
\label{eqn:4.4}%ラベル指定
\end{eqnarray}
{\bf 1.} \quad This term is completely written by the symmetric polynomials $S_i(z_\alpha)$ of the branch points, and can be identified as a holomorphic ambiguity of the BCOV holomorphic anomaly equation \cite{BCOV1,BKMP2}. Furthermore this term is important for deformations of the open string moduli, which correspond to the framing change of the A-brane, the change of the brane location and so on. These deformations are carried out by reparametrizations of the mirror curve $\Sigma$, which preserve the symplectic form $|dx/x \wedge dy/y|$ on $\Sigma$ \cite{BKMP1},
\begin{equation}
\begin{array}{ccccc}
PSL(2,{\IZ})~~~&  \curvearrowright \quad &\Sigma& \quad \longrightarrow \quad &\Sigma~~~ \\
\rotatebox{90}{$\in$}~~ &  & \rotatebox{90}{$\in$} &  &\rotatebox{90}{$\in$}~~~ \\
~\left(\begin{array}{cc} a & b \\ c & d \end{array}\right)~: &  & \left(\begin{array}{c} x \\ y \end{array}\right) & \quad \longmapsto \quad & \left(\begin{array}{c} x^ay^b \\ x^cy^d \end{array}\right)~.
\end{array}
\label{eqn:4.5}%ラベル指定
\end{equation}
Note that because the Bergman kernel defined as (\ref{eqn:2.4}) is irrelevant for this reparametrization of $\Sigma$, we can compute the Bergman kernel in another open string phase by a simple reparametrization of the open string moduli. In appendix C we will discuss the framing of the A-brane by this deformation.\\
{\bf 2.} \quad This term is important to discuss deformation of the closed string moduli which is carried out by the symplectic transformation of a canonical basis $(A,B)$ with non-trivial monodromy. By this transformation, in the K\"ahler moduli space of a local toric surface, we can move from the large radius phase to the orbifold phase via the mirror symmetry. Because in the following claim we show that this term can be written as a functional of the period, this symplectic transformation is carried out by a simple change of variables of the PF equations. In section 4.3 and 4.4 we will consider this transformation for several concrete examples.

Now we will compute the Bergman kernel from (\ref{eqn:2.13}) for the above $11$ local toric surfaces. For this purpose we should compute $G(\tau)$, and we prove the following formula as an extension of \cite{BKMP2}.\\
$\underline{\mbox{{\bf Claim :}}}$ \quad For the above $11$ local toric surfaces, by a change of variables in the complex moduli space ${\cal M}_B$ of a mirror curve $\Sigma$, $G(\tau)$ is given by
\begin{equation}
G(\tau)=-\frac{F(z_{\alpha})}{C_{zzz}}\theta\left\{12\log b(z_{\alpha})\theta T_z(z_{\alpha})+\log \Delta(z_{\alpha})\right\}~,\quad \theta:=z \frac{\partial}{\partial z}~,~F(z_{\alpha}):=\left(3ab(z_{\alpha})^2z^3\right)^{-1}~,
\label{eqn:4.6}%ラベル指定
\end{equation}
where $\Delta(z_{\alpha})$ is the discriminant of $\Sigma$ and ${C_{zzz}}$ is a Yukawa coupling, and $a$ and $b(z_{\alpha})$ are obtained in (\ref{eqn:4.9}) and (\ref{eqn:4.14}) (or (\ref{eqn:2.6})). $T_z(z_{\alpha})$ is a flat coordinate on ${\cal M}_B$ given by (\ref{eqn:3.14}) which has the form
\begin{equation}
T_z(z_{\alpha})=\log z+M(z_{\alpha}),\quad M(z_{\alpha}):~\mbox{a power series of}~z_1,\cdots,z_n~,
\label{eqn:4.7}%ラベル指定
\end{equation}
where we identify the parameter $z$ with $z_1$.\\
\noindent{\bf Proof)} \quad Let us consider the mirror variety $M_B$ of one of the above $11$ local toric surfaces with $n$ parameters, and make a change of variables in the complex moduli space ${\cal M}_B$ of $M_B$ such that the flat coordinates have the form
\begin{eqnarray}
T_z(z_{\alpha})&=&\log z+M(z_{\alpha})~,\quad \alpha=1,\cdots,n~,~z_1=z~,~\nonumber\\
T_{{\tilde \alpha}}(z_{\tilde \alpha})&=&\log z_{{\tilde \alpha}}+N(z_{\tilde \alpha})~,\quad {\tilde \alpha}=2,\cdots,n~,\quad N(z_{\tilde \alpha}):~\mbox{a power series of}~z_{\tilde \alpha}~\mbox{only}~.
\label{eqn:4.8}%ラベル指定
\end{eqnarray}
This change of variables is possible in our examples (see (\ref{eqn:4.32})), and in these variables $\tau$ is given by
\begin{equation}
\tau(z_{\alpha})=a \frac{\partial^2 {\cal F}_A^{(0,0)}}{\partial T_z^2}~,
\label{eqn:4.9}%ラベル指定
\end{equation}
where $a$ is a constant. As seen from (\ref{eqn:2.15}), $G(\tau)$ is constructed from two parts, $E_2(\tau)$ and $\omega_1$. At first let us consider $E_2(\tau)$. In lemma $1$ we will prove
\begin{equation}
E_2(\tau)=\frac{d}{d\tau}(12 \log \omega_1+\log \Delta(z_{\alpha}))=:\frac{d}{d\tau} f\left(\tau(z_{\alpha})\right)~.
\label{eqn:4.10}%ラベル指定
\end{equation}
From $\frac{\partial}{\partial z}f(\tau)=\frac{\partial \tau}{\partial z}\frac{d}{d \tau}f(\tau)$, we obtain
\begin{equation}
E_2(\tau)=\frac{d}{d\tau} f(\tau)=\left(\frac{\partial \tau}{\partial z}\right)^{-1}\frac{\partial}{\partial z}f(\tau)~.
\label{eqn:4.11}%ラベル指定
\end{equation}
By making use of (\ref{eqn:4.9}) and the Yukawa couplings $C_{\alpha\beta\gamma}\in Sym^3\left(T{\cal M}_B\right) \otimes {\cal L}^{-2}$, \footnote{${\cal L}$ is a holomorphic line bundle over $M_B$ such that the section is given by the holomorphic three-form $\Omega$ as (\ref{eqn:3.12}).}
\begin{eqnarray}
\frac{\partial \tau}{\partial z}&=& a \sum_{\delta=1}^n \frac{\partial T_{\delta}}{\partial z}\frac{\partial^3 {\cal F}_A^{(0,0)}}{\partial T_{\delta}\partial T_z^2}=a \sum_{\delta=1}^n \frac{\partial T_{\delta}}{\partial z} C_{T_{\delta}T_zT_z} \nonumber\\%~,\quad C_{T_{\delta}T_zT_z} \in Sym^3\left(T^*{\cal M}_A\right) \otimes {\cal L}^{-2}
&=& a \sum_{\alpha,\beta,\gamma,\delta=1}^n \frac{\partial T_{\delta}}{\partial z} \frac{\partial z_{\alpha}}{\partial T_{\delta}} \frac{\partial z_{\beta}}{\partial T_z} \frac{\partial z_{\gamma}}{\partial T_z} C_{\alpha\beta\gamma} = a \sum_{\beta,\gamma=1}^n \frac{\partial z_{\beta}}{\partial T_z} \frac{\partial z_{\gamma}}{\partial T_z} C_{z \beta\gamma} \nonumber\\
&=&a \left(\frac{\partial T_z}{\partial z}\right)^{-2} C_{zzz}~,
\label{eqn:4.12}%ラベル指定
\end{eqnarray}
where in the fourth and fifth equality, we used
$$
\delta_z^{\alpha}=\frac{\partial z_{\alpha}}{\partial z}=\sum_{\beta=1}^n \frac{\partial T_{\beta}}{\partial z}\frac{\partial z_{\alpha}}{\partial T_{\beta}}=\frac{\partial T_z}{\partial z}\frac{\partial z_{\alpha}}{\partial T_z} \quad \Longleftrightarrow \quad \frac{\partial z_{\alpha}}{\partial T_z}=\delta_z^{\alpha}\left(\frac{\partial T_z}{\partial z}\right)^{-1}~,
$$
as seen from our variables (\ref{eqn:4.8}). So from (\ref{eqn:4.11}) we obtain
\begin{equation}
E_2(\tau)=a^{-1} \left(\frac{\partial T_z}{\partial z}\right)^2 C_{zzz}^{-1} \frac{\partial}{\partial z} (12 \log \omega_1+\log \Delta(z_{\alpha}))~.
\label{eqn:4.13}%ラベル指定
\end{equation}
Next, let us consider $\omega_1$ and in lemma $2$, we will prove
\begin{equation}
\omega_1= i b(z_{\alpha}) \theta T_z~,\quad b(z_{\alpha}):~\mbox{a rational function of $z_{\alpha}$'s defined in (\ref{eqn:2.6})}~.
\label{eqn:4.14}%ラベル指定
\end{equation}
Therefore by combining (\ref{eqn:4.13}) and (\ref{eqn:4.14}), from (\ref{eqn:2.15}) we get the formula (\ref{eqn:4.6}). \hfill \qed \\%Claim QED
To complete our claim we show (\ref{eqn:4.10}) and (\ref{eqn:4.14}) in the following lemmas.\\
$\underline{\mbox{{\bf Lemma 1:}}}$ \quad $E_2(\tau)$ has the form (\ref{eqn:4.10}) for elliptic curves with genus one.\\
{\bf Proof)} \quad Firstly we note the Thomae's formula for a genus one Riemann surface (Prop. $3.6$ of \cite{Fay}),
\begin{eqnarray}
&&\vartheta_0^4=J(z_{\alpha})^2(s_1-s_2)(s_3-s_4)~,\nonumber\\
&&\vartheta_2^4=J(z_{\alpha})^2(s_1-s_4)(s_2-s_3)~,\nonumber\\
&&\vartheta_3^4=J(z_{\alpha})^2(s_1-s_3)(s_2-s_4)~,\nonumber\\
&&J(z_{\alpha}):=\oint_A \frac{dx}{\sqrt{\sigma(x)}}~,
\label{eqn:4.16}%ラベル指定
\end{eqnarray}
where $s_i$'s are the branch points of the mirror curve defined in (\ref{eqn:2.7}), and $A$ is an $A$-period on the Riemann surface. $\vartheta_0,~\vartheta_2$ and $\vartheta_3$ are the theta constants with $q=e^{2\pi i\tau}$ which are defined by
\begin{eqnarray}
\label{eqn:4.17}%ラベル指定
\vartheta_0:&=&1+2\sum_{n=1}^{\infty}(-1)^nq^{\frac{n^2}{2}}=q_0\prod_{n=1}^{\infty}\left(1-q^{\frac{2n-1}{2}}\right)^2~,\quad q_0:=\prod_{n=1}^{\infty}\left(1-q^n\right)~,\\
\label{eqn:4.18}%ラベル指定
\vartheta_2:&=&2\sum_{n=1}^{\infty}q^{\frac12(n-\frac12)^2}=2q^{\frac18}q_0\prod_{n=1}^{\infty}\left(1+q^n\right)^2~,\\
\vartheta_3:&=&1+2\sum_{n=1}^{\infty}q^{\frac{n^2}{2}}=q_0\prod_{n=1}^{\infty}\left(1+q^{\frac{2n-1}{2}}\right)^2~.
\label{eqn:4.19}%ラベル指定
\end{eqnarray}
Because the complete elliptic integral $K(k)$ is related to $\vartheta_3$ by
\begin{equation}
K(k)=\frac{\pi}{2}\vartheta_3^2~,
\label{eqn:4.20}%ラベル指定
\end{equation}
from (\ref{eqn:4.16}) we get
\begin{equation}
\vartheta_0^4\vartheta_2^4\vartheta_3^4=J(z_{\alpha})^6\Delta(z_{\alpha})^{\frac12}~,\quad \omega_1=\frac{2i}{\pi}\frac{K(k)}{\sqrt{(s_1-s_3)(s_2-s_4)}}=iJ(z_{\alpha})~.
\label{eqn:4.21}%ラベル指定
\end{equation}
On the other hand the product of $\vartheta_0,~\vartheta_2$ and $\vartheta_3$ is related to the Dedekind eta function by
\begin{equation}
\vartheta_0\vartheta_2\vartheta_3=2\eta(\tau)^3~,\quad \eta(\tau):=q^{\frac{1}{24}}\prod_{n=1}^{\infty}\left(1-q^n \right)=q^{\frac{1}{24}}q_0~.
\label{eqn:4.22}%ラベル指定
\end{equation}
From (\ref{eqn:4.21}) and (\ref{eqn:4.22}), we obtain
\begin{equation}
\log \eta(\tau)=\frac12\log \omega_1+\frac{1}{24}\Delta(z_{\alpha})+c~,
\label{eqn:4.23}%ラベル指定
\end{equation}
where $c$ is a constant. In the end, by making use of the formula
\begin{equation}
E_2(\tau)=24\frac{d}{d\tau}\log \eta(\tau)~,
\label{eqn:4.24}%ラベル指定
\end{equation}
we obtain (\ref{eqn:4.10}). \hfill \qed \\%Lemma QED
$\underline{\mbox{{\bf Lemma 2:}}}$ \quad $\omega_1$ is given by the form (\ref{eqn:4.14}) for the above $11$ local toric surfaces.\\
{\bf Proof)} \quad To prove this, it is enough that we prove this for $K_{dP_5}$ because others are obtained from $K_{dP_5}$ by blow-downs. For simplicity, we reparametrize the curve (\ref{eqn:4.2}) as $x \to x^{-1},~y \to x^{-2}y$ by making use of (\ref{eqn:4.5}),
\begin{equation}
{\widetilde \Sigma}~:\quad \lambda y^2-(x^2-\lambda x+{\tilde z}_5)y+\lambda^{-1}\sum_{\alpha=0}^4 {\tilde z}_{\alpha}(-x)^{4-\alpha}=0~.
\label{eqn:4.25}%ラベル指定
\end{equation}
For this curve, (\ref{eqn:2.7}) is obtained as
\begin{equation}
\sigma(x)=\frac{1}{1-4{\tilde z}_0}\left\{(x^2-\lambda x+{\tilde z}_5)^2-4\sum_{\alpha=0}^4 {\tilde z}_{\alpha}(-x)^{4-\alpha}\right\}=:\frac{1}{1-4{\tilde z}_0}{\widetilde \sigma}(x)~.
\label{eqn:4.26}%ラベル指定
\end{equation}
From (\ref{eqn:4.21}), we rewrite $\omega_1$ by $\omega(x)$ in (\ref{eqn:3.12}) as \cite{BrTan}
\begin{eqnarray}
\omega_1 &=& i \oint_{A}\frac{dx}{\sqrt{\sigma(x)}} = 
i \left(1-4{\tilde z}_0\right)^{\frac12} \oint_{A}\frac{dx}{\sqrt{{\widetilde \sigma}(x)}} \nonumber\\
&=& -i \left(1-4{\tilde z}_0\right)^{\frac12} \frac{\partial}{\partial \lambda} \oint_{A} \log \left(\frac{(x^2-\lambda x+{\tilde z}_5)+\sqrt{{\widetilde \sigma}(x)}}{2}\right)\frac{dx}{x} \nonumber\\
&=& -i \left(1-4{\tilde z}_0\right)^{\frac12} \frac{\partial}{\partial \lambda} \oint_{A} \omega(x)~.
\label{eqn:4.27}%ラベル指定
\end{eqnarray}
Here we note
\begin{equation}
-\lambda \frac{\partial}{\partial \lambda}=\sum_{\alpha=1}^5D_{\alpha}\theta_{\alpha}~,\quad \theta_{\alpha}=z_{\alpha}\frac{\partial}{\partial z_{\alpha}},~D_{\alpha}:=\alpha,~\mbox{for}~\alpha=0,\cdots,4~\mbox{and}~ D_5:=2~.
\label{eqn:4.28}%ラベル指定
\end{equation}
On the other hand, we have to make a change of variables such that the flat coordinates have the same form as (\ref{eqn:4.8}). For charge vectors (\ref{eqn:4.1}), from (\ref{eqn:3.17}) we obtain the flat coordinates $T_{\alpha}$ as follows \cite{LerM} ;
\begin{eqnarray}
\label{eqn:4.29}%ラベル指定
T_{\alpha}&&=\log z_{\alpha}+\alpha M(z)+(2-\alpha)N(z_0)~,~\alpha=0,\cdots,4~,\quad T_5=\log z_5+2 M(z)-N(z_0)~,\\
&&M(z):=\sum_{m_{\alpha}\geq 0,~(\alpha-2)m_{\alpha}+m_5\geq 0}\frac{(-1)^{m_1+m_3}\left(\sum_{\alpha=0}^4\alpha m_{\alpha}+2m_5-1\right)!}{\left(\sum_{\alpha=0}^4m_{\alpha}\right)!\left(\sum_{\alpha=0}^4(\alpha-2)m_{\alpha}+m_5\right)!\prod_{\alpha=0}^5m_{\alpha}!}\prod_{\alpha=0}^5z_{\alpha}^{m_{\alpha}}~, \hspace{3em}\nonumber\\
&&N(z_0):=\sum_{m=1}^{\infty}\frac{(2m-1)!}{(m!)^2}z_0^m=-\log\frac{1+\sqrt{1-4z_0}}{2}~.\nonumber
\end{eqnarray}
Now let us define 
\begin{eqnarray}
\label{eqn:4.30}%ラベル指定
&&{\overline T}_0=T_0~,\quad {\overline T}_{\alpha}=T_{\alpha}-\frac{\alpha}{2}T_5-\left(1-\frac{\alpha}{4}\right)T_0~,~\alpha=1,\cdots,4~,\quad {\overline T}_5=T_5+\frac12 T_0=:T_z~,\\
&&{\overline z}_0=z_0~,\quad {\overline z}_{\alpha}=z_{\alpha}z_5^{-\frac{\alpha}{2}}z_0^{-\left(1-\frac{\alpha}{4}\right)}~,\quad {\overline z}_5=z_0^{\frac12}z_5=:z~.
\label{eqn:4.31}%ラベル指定
\end{eqnarray}
In these new variables, the flat coordinates (\ref{eqn:4.29}) are rewritten as 
\begin{equation}
{\overline T}_0=\log {\overline z}_0+2N({\overline z}_0)~,\quad {\overline T}_{\alpha}=\log {\overline z}_{\alpha}~,\quad {\overline T}_5=\log {\overline z}_5+2M({\overline z})~.
\label{eqn:4.32}%ラベル指定
\end{equation}
Because we obtain
$$
\sum_{\alpha=1}^5D_{\alpha}\theta_{\alpha}=\sum_{\alpha,\beta=1}^5D_{\alpha}z_{\alpha}\frac{\partial {\overline z}_{\beta}}{\partial z_{\alpha}}\frac{\partial}{\partial {\overline z}_{\beta}}=2z\frac{\partial}{\partial z}=2\theta~,
$$
from (\ref{eqn:4.28}), by normalizing to $\lambda=1$ we can rewrite (\ref{eqn:4.27}) as
\begin{equation}
\omega_1=2i \left(1-4{\overline z}_0\right)^{\frac12}\theta \oint_{A} \omega(x)~.
\label{eqn:4.33}%ラベル指定
\end{equation}
From (\ref{eqn:3.14}), we can choose a cycle $A_z$ such as
$$
T_z({\overline z}_{\alpha})=\oint_{A_z}\omega(x)~,
$$
and from the asymptotic behavior $\omega_1 \sim i+{\cal O}({\overline z}_{\alpha})$, we find that two cycles $A$ and $A_z$ should be related by $2A=A_z$ up to integrals around punctures, so we obtain
\begin{equation}
\omega_1=i\left(1-4{\overline z}_0\right)^{\frac12} \theta T_z({\overline z}_{\alpha})~.
\label{eqn:4.34}%ラベル指定
\end{equation}
By $b({\overline z}_{\alpha})=(1-4{\overline z}_0)^{1/2}$, our lemma is proved. Note that in the case of $z_5=0$, we can also make a change of variables such as (\ref{eqn:4.30}) and (\ref{eqn:4.31}). In general, from (\ref{eqn:4.27}) we see that $b(z_{\alpha})$ is given in (\ref{eqn:2.6}).  \hfill \qed \\%Lemma QED

\subsection{Examples}%sec4.3

\noindent Here we give several concrete examples of the annulus amplitude. In the following examples, for simplicity we reparametrize the mirror curve as $x \to x^{-1},~y \to x^{-2}y$, and put $\lambda=1$. In this open string phase, by a change of variables (\ref{eqn:4.31}) the mirror curve of $K_{dP_5}$ is given by
\begin{equation}
{\widetilde \Sigma}~:\quad y^2-(x^2-x+z_0^{-\frac12}z)y+z_0x^4+\sum_{\alpha=1}^4 z^{\frac{\alpha}{2}}z_0^{1-\frac{\alpha}{2}}z_{\alpha}(-x)^{4-\alpha}=0~,
\label{eqn:4.35}%ラベル指定
\end{equation}
where we abbreviated ${\overline z}_{\alpha}$ to $z_{\alpha}$. From (\ref{eqn:4.32}) the closed mirror maps are given by 
\begin{equation}
T_0=\log z_0+2N(z_0)~,\quad T_{\alpha}=\log z_{\alpha}~,~\alpha=1,\cdots,4~,\quad T_z=\log z+2M(z_{\alpha})~.
\label{eqn:4.36}%ラベル指定
\end{equation}
Taking account of the above reparametrization, we obtain the annulus amplitude from (\ref{eqn:4.6}),
\begin{eqnarray}
\label{eqn:4.37}%ラベル指定
&&A_B(x_1,x_2):={\cal F}_B^{(0,2)}(x_1,x_2;z_{\alpha})=F^{(0,2)}(x_1,x_2;z_{\alpha})=\int^{x_1}\int^{x_2} W^{(0,2)}(x_1^{-1},x_2^{-1})\frac{1}{x_1^2x_2^2}~,\\
&&\quad W^{(0,2)}(x_1,x_2)=\frac{dx_1dx_2}{\sqrt{\sigma(x_1)\sigma(x_2)}}\left(\frac{-\sqrt{\sigma(x_1)\sigma(x_2)}+f(x_1,x_2)}{2(x_1-x_2)^2}+\frac{G(\tau)}{4}\right)~,\hspace{3em}\nonumber\\
&&\quad G(\tau)=-\frac{F(z_{\alpha})}{C_{zzz}}\theta\left\{12\log(1-4z_0)^{\frac12}\theta T_z(z_{\alpha})+\log \Delta(z_{\alpha})\right\}~,\quad F(z_{\alpha})=\left(3a(1-4z_0)z^3\right)^{-1}~,\nonumber
\end{eqnarray}
where in the second equality of (\ref{eqn:4.37}), we used the BKMP conjecture (\ref{eqn:3.13}). Actually in the following examples we can check this conjecture. The open mirror map (\ref{eqn:3.16}) is given by \cite{LerM}
\begin{equation}
X=-e^{-M(z_{\alpha})}x=-\left(\frac{z}{Q_z}\right)^{\frac12}x~,\quad Q_z:=e^{T_z}~.
\label{eqn:4.38}%ラベル指定
\end{equation}
By applying the mirror maps (\ref{eqn:4.36}) and (\ref{eqn:4.38}) to (\ref{eqn:4.37}), we can obtain the annulus amplitude $A_A(X_1,X_2)$ on $K_{dP_5}$. Note that for $9$ local toric surfaces of $11$ local toric surfaces, we can use above strategy, whereas for $2$ local toric del Pezzo surfaces $K_{{\IP}^2}$ and $K_{dP_3}$ (corresponding to no.$1$ and no.$10$ of figure \ref{fig:1}) we cannot use above computation because $z_5=0$. But the same way as the change of variables (\ref{eqn:4.30}) and (\ref{eqn:4.31}), we can obtain the annulus amplitudes on these varieties. As examples, let us consider $K_{{\IP}^2},~K_{{\IF}_0}$, $K_{dP_2^{(0)}}$ and $K_{{\IF}_2}$ (no.1, no.2, no.5 and no.4 of figure \ref{fig:1}). These examples were investigated in \cite{Mar1,BKMP1,BKMP2,BrTan}, and recently based on the BCOV holomorphic anomaly equation, the closed string higher amplitudes were computed for these examples in \cite{HKR,ALM}. Because $K_{{\IP}^2}$ and its orbifold phase ${\IC}^3/{\IZ}_3$ were studied in \cite{Mar1,BKMP1} and \cite{ABK,BKMP2}, at first we consider the annulus amplitude on $K_{{\IF}_0}$ based on the formula (\ref{eqn:4.6}).\\
$\underline{\mbox{{\bf Example 1-1:}}~K_{{\IF}_0}}$ \cite{Mar1}

\begin{wrapfigure}{r}{35mm}
 \begin{center}
  \includegraphics[width=23mm]{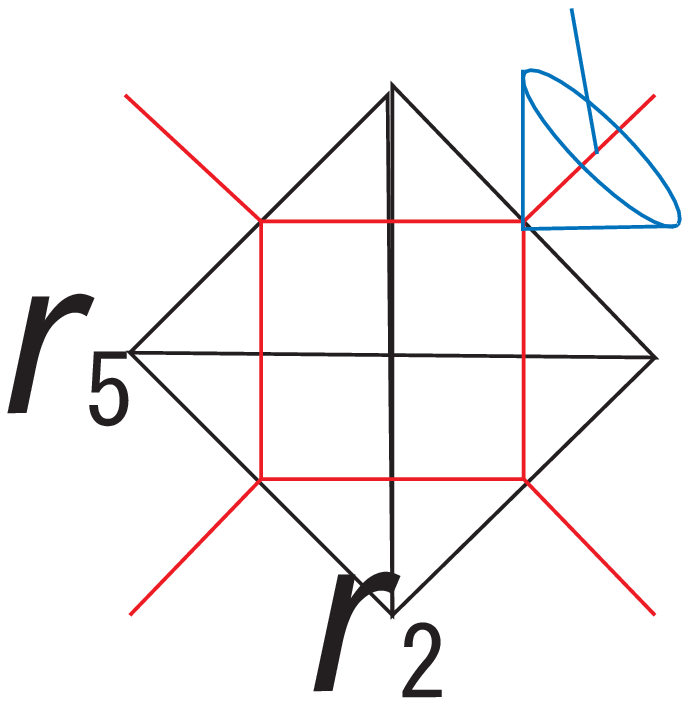}
  \caption{Fan of $K_{{\IF}_0}$: $r_2$ and $r_5$ are the K\"ahler parameters defined in (\ref{eqn:3.10}).}
  \label{fig:2}
 \end{center}
\end{wrapfigure}

\noindent Because the charge vectors of $K_{{\IF}_0}$ are given by
$$
l^2=(-2,1,0,1,0)~,\quad l^5=(-2,0,1,0,1)~,
$$
after a reparametrization $x \to x^{-1},~y \to x^{-2}y$, we obtain a mirror curve of this variety
\begin{equation}
y^2-(x^2-x+z)y+zz_2x^2=0~,\quad \sigma(x)=(x^2-x+z)^2-4zz_2x^2=\prod_{i=1}^4(x-s_i)~.
\label{eqn:4.39}%ラベル指定
\end{equation}
The inserted A-brane represented by the blue line in figure \ref{fig:2} is well described by this curve. The necessary data to compute $G(\tau)$ from (\ref{eqn:4.37}) are
\begin{eqnarray}
&&\Delta(z_{\alpha})=\prod_{i=1}^4\sigma '(s_i)=4^4z^4z_2^2\Delta_0(z_{\alpha})~,~\Delta_0(z_{\alpha}):=1-8z(1+z_2)+16z^2(1-z_2)^2~, \nonumber\\
&&C_{zzz}=-\frac{1}{z^3\Delta_0(z_{\alpha})}~,\quad \mbox{and}~a=-4~,~b=1~,~\mbox{so}\quad F(z_{\alpha})=\frac{-1}{12z^3}~,
\label{eqn:4.40}%ラベル指定
\end{eqnarray}
where the Yukawa coupling and $a=-4$ has been obtained in \cite{FoJi} and \cite{HKR}. From this data, we obtain
\begin{equation}
G(\tau)=-\frac{1}{12}\Delta_0(z_{\alpha})\theta \left\{12\log \theta T_z(z_{\alpha})+4\log z +\log \Delta_0(z_{\alpha})\right\}~.
\label{eqn:4.41}%ラベル指定
\end{equation}
The closed and open mirror maps are given by
\begin{equation}
T_z(z_{\alpha})=\log z+2\sum_{m,n\ge 0}^{\infty}\frac{(2m+2n-1)!}{m!^2n!^2}z^{m+n}z_2^n~,\quad T_2(z_2)=\log z_2~,
\label{eqn:4.42}%ラベル指定
\end{equation}
and (\ref{eqn:4.38}) respectively. By (\ref{eqn:4.30}), we write in the original variables $Q_2$ and $Q_5$,
$$
Q_2~\longrightarrow~\frac{Q_2}{Q_5}~,\quad Q_z~\longrightarrow~Q_5~.
$$
From these, we obtain the annulus amplitude
\begin{equation}
A_A(X_1,X_2)=\int^{x_1}\int^{x_2} W^{(0,2)}(x_1^{-1},x_2^{-1})\frac{1}{x_1^2x_2^2} = \sum_{i_1,i_2}A^{(i_1,i_2)}(\frac{Q_2}{Q_5},Q_5)X_1^{i_1}X_2^{i_2}~,
\label{eqn:4.43}%ラベル指定
\end{equation}
where $A^{(i_1,i_2)}(Q_{\alpha})=A^{(i_1,i_2)}(Q_5,Q_2/Q_5)$ are
\begin{eqnarray}
&&A^{(1,1)}(Q_{\alpha})= 
Q_2+2Q_2Q_5+4Q_2Q_5(Q_2+Q_5)+6Q_2Q_5(Q_2^2+6Q_2Q_5+Q_5^2) \nonumber\\
&&\quad +8Q_2Q_5(Q_2^3+20Q_2^2Q_5+20Q_2Q_5^2+Q_5^3)+10Q_2Q_5(Q_2^4+50Q_2^3Q_5+155Q_2^2Q_5^2+\cdots)+\cdots~,\nonumber\\
&&A^{(1,2)}(Q_{\alpha})= 
-Q_2-Q_2(Q_2+2Q_5)-3Q_2Q_5(2Q_2+Q_5)-5Q_2Q_5(2Q_2^2+6Q_2Q_5+Q_5^2) \nonumber\\
&&\quad -7Q_2Q_5(2Q_2^3+21Q_2^2Q_5+17Q_2Q_5^2+Q_5^3)-9Q_2Q_5(2Q_2^4+55Q_2^3Q_5+131Q_2^2Q_5^2+\cdots)+\cdots~,\nonumber\\
&&A^{(1,3)}(Q_{\alpha})= Q_2+Q_2(3Q_2+2Q_5)+Q_2(Q_2^2+12Q_2Q_5+3Q_5^2)+4Q_2Q_5(6Q_2^2+10Q_2Q_5+Q_5^2) \nonumber\\
&&\quad +3Q_2Q_5(12Q_2^3+75Q_2^2Q_5+40Q_2Q_5^2+2Q_5^3)+8Q_2Q_5(6Q_2^4+104Q_2^3Q_5+172Q_2^2Q_5^2+\cdots)+\cdots~,\nonumber\\
&&A^{(2,2)}(Q_{\alpha})= Q_2+Q_2(\frac52 Q_2+2Q_5)+Q_2(Q_2^2+10Q_2Q_5+3Q_5^2)+Q_2Q_5(20Q_2^2+33Q_2Q_5+4Q_5) \nonumber\\
&&\quad +3Q_2Q_5(10Q_2^3+59Q_2^2Q_5+34Q_2Q_5^2+2Q_5^3)+2Q_2Q_5(20Q_2^4+321Q_2^3Q_5+544Q_2^2Q_5^2+\cdots)+\cdots~.\nonumber
\end{eqnarray}
These amplitudes agree with the topological vertex calculus with the framing ${\widetilde f}=-1$ in appendix A as far as we have checked.\\
$\underline{\mbox{{\bf Example 1-2:}}~T^*S^3/{{\IZ}_2}}$ \cite{AKMV,BKMP1}\\
Next we consider the orbifold phase $T^*S^3/{{\IZ}_2}$ of $K_{{\IF}_0}$ considered in \cite{AKMV}, and further investigated in \cite{BKMP1}. Following \cite{AKMV}, we consider a transformation $z=(q_1q_2)^{-2},~z_2=1-q_1$ of the closed string moduli. In these new coordinates $q_1$ and $q_2$, by considering the PF operators (\ref{eqn:3.17}),
\begin{eqnarray}
{\cal D}_2&=&q_2^2\left\{(1-q_1)(\theta_1-\theta_2)-1\right\}(\theta_1-\theta_2)-\theta_2(\theta_2-1)~,\quad \theta_i:=q_i\frac{\partial}{\partial q_i}~,\nonumber\\
{\cal D}_5&=&q_2^2\left\{2(1-q_1)\theta_1-(2-q_1)\theta_2-2(1-q_1)\right\}\left\{2(1-q_1)\theta_1-(2-q_1)\theta_2\right\}-4\theta_2(\theta_2-1)~,\nonumber
\end{eqnarray}
we obtain the orbifold mirror maps
\begin{eqnarray}
s_1(q_1)&=& -\log(1-q_1)~,\nonumber\\
s_2(q_{\alpha})&=& \sum_{n=0,m=2n+1}^{\infty}\frac{(2n)!(2n-1)!!\left[(2m-2n-3)!!\right]^2}{4^{m+n-1}n!^2(m-1)!(m-2n-1)!(2n+1)!!}q_1^mq_2^{2n+1}~.
\label{eqn:4.44}%ラベル指定
\end{eqnarray}
From (\ref{eqn:4.41}), we can obtain $G(\tau)$ in this orbifold phase by replacing $T_z(z_{\alpha})$ with $s_2(q_{\alpha})$,
\begin{eqnarray}
&&\Delta_0(z_{\alpha})=(q_1q_2)^{-4}\left((q_1q_2)^4-8(q_1q_2)^2(2-q_1)+16q_1^2\right)=:(q_1q_2)^{-4}{\widetilde \Delta_0(q_{\alpha})}~,\hspace{10em}\nonumber\\
&&G(\tau)=\frac{1}{24}(q_1q_2)^{-4}{\widetilde \Delta_0(q_{\alpha})}\theta_{q_2} \left\{12\log \theta_{q_2} s_2(q_{\alpha})-12\log q_2 +\log {\widetilde \Delta_0(q_{\alpha})}\right\}~,\quad \theta_{q_2}:=q_2\frac{\partial}{\partial q_2}~.
\label{eqn:4.45}%ラベル指定
\end{eqnarray}
The open orbifold mirror map was suggested in \cite{BKMP1}, and in this case this is given by
\begin{equation}
X=-q_1(s_1)^{-1}q_2(s_1,s_2)^{-1}x~,
\label{eqn:4.46}%ラベル指定
\end{equation}
and let us transform the closed string variables $s_1$ and $s_2$ to the Chern-Simons matrix model variables $S_1=(s_1+s_2)/4$ and $S_2=(s_1-s_2)/4$. Then we obtain the orbifold annulus amplitude $A_o(X_1,X_2)$ on $T^*S^3/{{\IZ}_2}$,
\begin{equation}
A_o(X_1,X_2)=\int^{x_1}\int^{x_2} W_o^{(0,2)}(x_1^{-1},x_2^{-1})\frac{1}{x_1^2x_2^2} = \sum_{i_1,i_2}A_o^{(i_1,i_2)}(S_{\alpha})X_1^{i_1}X_2^{i_2}~,
\label{eqn:4.47}%ラベル指定
\end{equation}
where if $i_1+i_2=$even (odd), then $A_o^{(i_1,i_2)}(S_{\alpha})$ are symmetric (antisymmetric) under $S_1\leftrightarrow S_2$ as,
\begin{eqnarray}
&&A_o^{(1,1)}(S_{\alpha})= -(S_1+S_2)+\frac12(3S_1^2+4S_1S_2+3S_2^2)-\frac16(7S_1^3+15S_1^2S_2+\cdots) \nonumber\\
&&\quad +\frac{1}{24}(15S_1^4+47S_1^3S_2+63S_1^2S_2^2+\cdots)-\frac{1}{120}(31S_1^5+130S_1^4S_2+235S_1^3S_2^2+\cdots)+\cdots~,\nonumber\\
&&A_o^{(1,2)}(S_{\alpha})= -(S_1-S_2)\left\{1-\frac72(S_1+S_2)+\frac{1}{12}(62S_1^2+113S_1S_2+62S_2^2)\right. \nonumber\\
&&\quad \left.-\frac{1}{24}(115S_1^3+316S_1^2S_2+\cdots)+\frac{1}{3072}(10035S_1^4+37540S_1^3S_2+55010S_1^2S_2^2+\cdots)-\cdots\right\}~,\nonumber\\
&&A_o^{(1,3)}(S_{\alpha})= -(S_1+S_2)+\frac12(13S_1^2+4S_1S_2+13S_2^2)-\frac16(97S_1^3+57S_1^2S_2+\cdots) \nonumber\\
&&\quad +\frac{1}{128}(2985S_1^4+3724S_1^3S_2+1478S_1^2S_2^2+\cdots)-\frac{1}{1920}(45571S_1^5+93575S_1^4S_2+52870S_1^3S_2^2+\cdots)+\cdots~,\nonumber\\
&&A_o^{(2,2)}(S_{\alpha})= -(S_1+S_2)+6(S_1^2+S_2^2)-\frac43(11S_1^3+3S_1^2S_2+\cdots) \nonumber\\
&&\quad +\frac18(169S_1^4+160S_1^3S_2-18S_1^2S_2^2+\cdots)-\frac{1}{60}(3\cdot 431S_1^5+5\cdot 457S_1^4S_2+13\cdot 30S_1^3S_2^2+\cdots)+\cdots~.\nonumber
\end{eqnarray}
By $S_1 \to -S_1$ and $S_2 \to -S_2$, these amplitudes completely agree with the computation of the Wilson loop in the Chern-Simons theory on $S^3/{\IZ}_2$ \cite{BKMP1}.\\
$\underline{\mbox{{\bf Example 2:}}~K_{dP_2^{(0)}}}$\vspace{0.3em}

\begin{wrapfigure}{r}{35mm}
 \begin{center}
  \includegraphics[width=30mm]{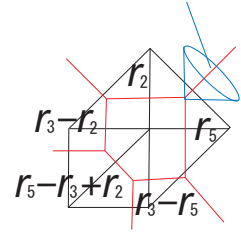}
  \caption{Fan of $K_{dP_2^{(0)}}$: $r_2$, $r_3$ and $r_5$ are the K\"ahler parameters defined in (\ref{eqn:3.10}).}
  \label{fig:3.5}
 \end{center}
\end{wrapfigure}

\noindent As we see from figure \ref{fig:1}, the moduli space of $K_{dP_2^{(0)}}$ (no.5) contains the moduli spaces of the geometries of no.1, no.2 and no.3. Because the charge vectors of this geometry are given by
$$
l^2=(-2,1,0,1,0,0)~,\quad l^3=(-3,1,1,0,1,0)~,\quad l^5=(-2,0,1,0,0,1)~,
$$
after a reparametrization $x \to x^{-1},~y \to x^{-2}y$, we obtain a mirror curve
\begin{eqnarray}
&&y^2-(x^2-x+z)y+zz_2x^2-z^{\frac32}z_3x=0~,\nonumber\\
&&\sigma(x)=(x^2-x+z)^2-4z(z_2x-z^{\frac12}z_3)x=\prod_{i=1}^4(x-s_i)~,
\label{eqn:4.481}%ラベル指定
\end{eqnarray}
which describes the moduli space of the A-brane in figure \ref{fig:3.5}. Note that by taking a limit $z_3=0$, (or $z_2=0$, or $z_2=z=0$ under fixing to $z^{\frac32}z_3$), we obtain the mirror curve which describes the moduli space of the A-brane of no.2, (or no.3, or no.1) in figure \ref{fig:1}. The necessary data to compute $G(\tau)$ are
\begin{eqnarray}
&&\Delta(z_{\alpha})=\prod_{i=1}^4\sigma '(s_i)=4^4z^4z_2^2\Delta_0(z_{\alpha})~,\hspace{29.5em}\nonumber\\
&&\Delta_0(z_{\alpha}):=-27z^2z_2^{-2}z_3^4-z^{\frac12}z_2^{-2}\bigl\{1-36z(1+z_2)\bigr\}z_3^3+z_2^{-2}\bigl\{1+z_2-2z(4+23z_2+4z_2^2) \nonumber\\
&&\hspace{5em}+8z^2(2-3z_2-3z_2^2+2z_2^3)\bigr\}z_3^2 -z^{-\frac12}z_2^{-1}\bigl\{1-8z(1+z_2)+16z^2(1-4z_2+z_2^2)\bigr\}z_3 \nonumber\\
&&\hspace{5em}+\bigl\{1-8z(1+z_2)+16z^2(1-z_2)^2\bigr\}~,\nonumber\\
&&C_{zzz}=-\frac{D(z_{\alpha})}{z^3\Delta_0(z_{\alpha})}~,\quad D(z_{\alpha}):=-\frac{9}{8}z^{\frac12}z_2^{-2}z_3^3+(1+z_2)z_2^{-2}z_3^2-\frac{1}{8}(7+4z+4zz_2)z^{-\frac12}z_2^{-1}z_3+1~,
\label{eqn:4.491}%ラベル指定
\end{eqnarray}
where the Yukawa coupling has been obtained in \cite{FoJi}. We can also find $b=1$ from (\ref{eqn:4.481}), and $a=-4$ from the consistency with example 1-1 by $z_3=0$. By these data we obtain,
\begin{equation}
G(\tau)=-\frac{1}{12D(z_{\alpha})}\Delta_0(z_{\alpha})\theta \left\{12\log \theta T_z(z_{\alpha})+4\log z +\log \Delta_0(z_{\alpha})\right\}~.
\label{eqn:4.501}%ラベル指定
\end{equation}
The closed and open mirror maps are given by
\begin{equation}
T_z(z_{\alpha})=\log z+2\sum_{p,q,r\ge 0}^{\infty}\frac{(-1)^q(2p+3q+2r-1)!}{p!q!r!(p+q)!(q+r)!}z_2^pz_3^qz^{p+\frac32q+r}~,\quad T_2(z_2)=\log z_2~,\quad T_3(z_3)=\log z_3~,
\label{eqn:4.511}%ラベル指定
\end{equation}
and (\ref{eqn:4.38}) respectively. By (\ref{eqn:4.30}), we write in the original variables $Q_2$, $Q_3$ and $Q_5$,
$$
Q_2~\longrightarrow~\frac{Q_2}{Q_5}~,\quad Q_3~\longrightarrow~\frac{Q_3}{Q_5^{\frac32}}~,\quad Q_z~\longrightarrow~Q_5~.
$$
As a result, we obtain the annulus amplitude
\begin{equation}
A_A(X_1,X_2)=\int^{x_1}\int^{x_2} W^{(0,2)}(x_1^{-1},x_2^{-1})\frac{1}{x_1^2x_2^2} = 
\sum_{i_1,i_2}A^{(i_1,i_2)}(\frac{Q_2}{Q_5},\frac{Q_3}{Q_5^{\frac32}},Q_5)X_1^{i_1}X_2^{i_2}~,
\label{eqn:4.521}%ラベル指定
\end{equation}
where $A^{(i_1,i_2)}(Q_{\alpha})=A^{(i_1,i_2)}(Q_2/Q_5,Q_3/Q_5^{\frac32},Q_5)$ are
\begin{eqnarray}
A^{(1,1)}(Q_{\alpha})&=& 
Q_2-Q_3+2Q_2Q_5-3Q_3Q_5-3Q_2Q_3+4Q_3^2+4Q_2Q_5^2+4Q_2^2Q_5-5Q_3Q_5^2+\cdots~,\nonumber\\
A^{(1,2)}(Q_{\alpha})&=& 
-Q_2+Q_3-2Q_2Q_5+2Q_3Q_5+4Q_2Q_3-Q_2^2-3Q_3^2-3Q_2Q_5^2-6Q_2^2Q_5+4Q_3Q_5^2+\cdots~,\nonumber\\
A^{(1,3)}(Q_{\alpha})&=& Q_2-Q_3+2Q_2Q_5-2Q_3Q_5-7Q_2Q_3+3Q_2^2+4Q_3^2+3Q_2Q_5^2+12Q_2^2Q_5-3Q_3Q_5^2+\cdots~,\nonumber\\
A^{(2,2)}(Q_{\alpha})&=& Q_2-Q_3+2Q_2Q_5-2Q_3Q_5-6Q_2Q_3+\frac{5}{2}Q_2^2+\frac{7}{2}Q_3^2+3Q_2Q_5^2+10Q_2^2Q_5-3Q_3Q_5^2+\cdots~.\nonumber
\end{eqnarray}
These amplitudes agree with the topological vertex calculus with the framing ${\widetilde f}=-1$ in appendix A as far as we have checked. And by taking a limit $Q_3=0$, (or $Q_2=0$, or $Q_2=Q_5=0$), we obtain the annulus amplitude on $K_{{\IF}_0}$, (or $K_{{\IF}_1}$, or $K_{{\IP}^2}$).

\noindent$\underline{\mbox{{\bf Example 3:}}~K_{{\IF}_2}}$

\begin{wrapfigure}{r}{35mm}
 \begin{center}
  \includegraphics[width=28mm]{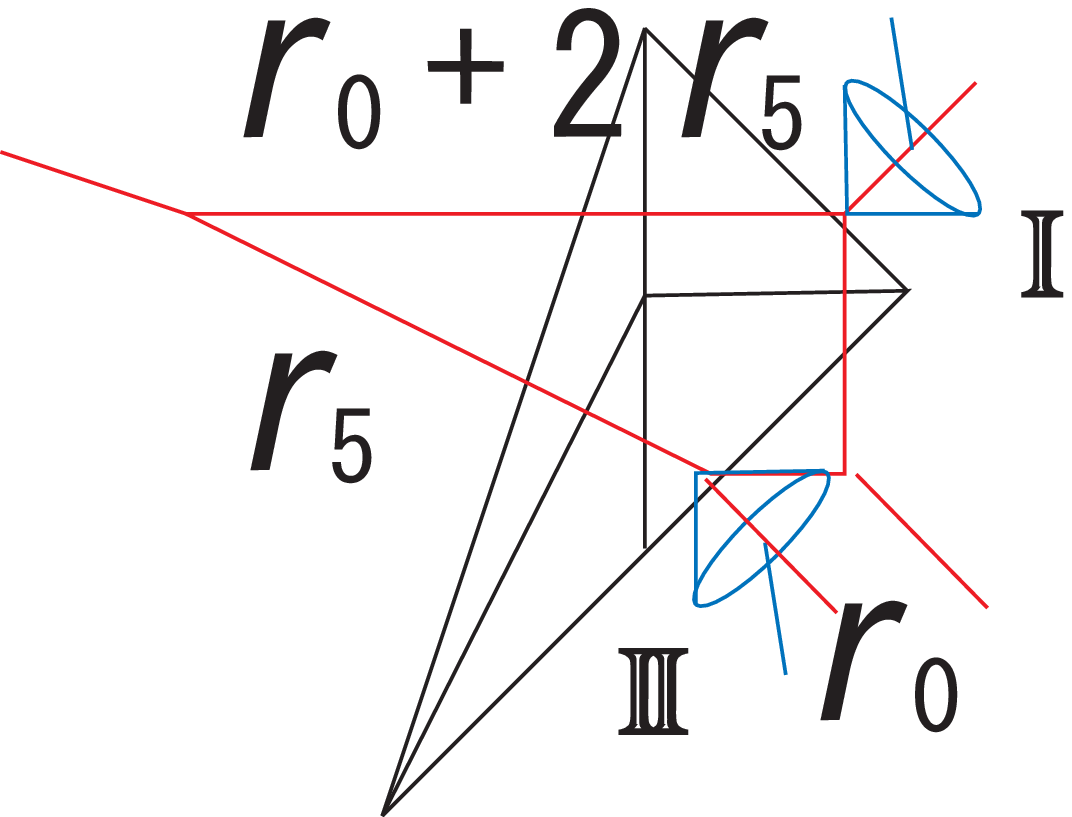}
  \caption{Fan of $K_{{\IF}_2}$: $r_0$ and $r_5$ are the K\"ahler parameters defined in (\ref{eqn:3.10}).}
  \label{fig:3}
 \end{center}
\end{wrapfigure}

\noindent Because the charge vectors of $K_{{\IF}_2}$ are given by
$$
l^0=(0,1,-2,1,0)~,\quad l^5=(-2,0,1,0,1)~,
$$
after a reparametrization $x \to x^{-1},~y \to x^{-2}y$, we obtain a mirror curve of this variety as (\ref{eqn:B.10}) which corresponds to the open string phase {\II} in figure \ref{fig:3}, but here we use another mirror curve,
\begin{equation}
y^2-(x^2-x+z_0^{-\frac12}z)y+z^2=0~,\quad \sigma(x)=(x^2-x+z_0^{-\frac12}z)^2-4z^2=\prod_{i=1}^4(x-s_i)~,
\label{eqn:4.48}%ラベル指定
\end{equation}
which corresponds to the open string phase I in figure \ref{fig:3}. The necessary data to compute $G(\tau)$ are
\begin{eqnarray}
&&\Delta(z_{\alpha})=\prod_{i=1}^4\sigma '(s_i)=4^4z^4\Delta_0(z_{\alpha})~,~\Delta_0(z_{\alpha}):=(1-4z_0^{-\frac12}z)^2-64z^2~, \nonumber\\
&&C_{zzz}=-\frac{1}{z^3\Delta_0(z_{\alpha})}~,\quad \mbox{and}~a=-4~,~b=1~,~\mbox{so}\quad F(z_{\alpha})=\frac{-1}{12z^3}~,
\label{eqn:4.49}%ラベル指定
\end{eqnarray}
where the Yukawa coupling has been obtained in \cite{FoJi} and $a=-4$ is fixed by the asymptotic behavior of the Bergmann kernel (\ref{eqn:2.13}). From this data, by (\ref{eqn:4.37}) we obtain the same form as (\ref{eqn:4.41}),
\begin{equation}
G(\tau)=-\frac{1}{12}\Delta_0(z_{\alpha})\theta \left\{12\log \theta T_z(z_{\alpha})+4\log z +\log \Delta_0(z_{\alpha})\right\}~.
\label{eqn:4.50}%ラベル指定
\end{equation}
The closed and open mirror maps are given by
\begin{equation}
T_z(z_{\alpha})=\log z+2\sum_{m,n \ge 0,~m \ge 2n}^{\infty}\frac{(2m-1)!}{m!n!^2(m-2n)!}z^mz_0^{-\frac{m}{2}+n}~,\quad T_0(z_0)=\log z_0-2\log \frac{1+\sqrt{1-4z_0}}{2}~,
\label{eqn:4.51}%ラベル指定
\end{equation}
and (\ref{eqn:4.38}) respectively. By (\ref{eqn:4.30}), we write in the original variables $Q_0$ and $Q_5$,
$$
Q_0~\longrightarrow~Q_0~,\quad Q_z~\longrightarrow~Q_0^{\frac12}Q_5~.
$$
As a result, we obtain the annulus amplitude
\begin{equation}
A_A(X_1,X_2)=\int^{x_1}\int^{x_2} W^{(0,2)}(x_1^{-1},x_2^{-1})\frac{1}{x_1^2x_2^2} = 
\sum_{i_1,i_2}A^{(i_1,i_2)}(Q_0,Q_0^{\frac12}Q_5)X_1^{i_1}X_2^{i_2}~,
\label{eqn:4.52}%ラベル指定
\end{equation}
where $A^{(i_1,i_2)}(Q_{\alpha})=A^{(i_1,i_2)}(Q_0,Q_0^{\frac12}Q_5)$ are
\begin{eqnarray}
A^{(1,1)}(Q_{\alpha})&=& 
2Q_0Q_5^2+4Q_0Q_5^3+6Q_0Q_5^4+4Q_0^2Q_5^3+8Q_0Q_5^5+36Q_0^2Q_5^4+160Q_0^2Q_5^5+6Q_0^3Q_5^4+\cdots~,\nonumber\\
A^{(1,2)}(Q_{\alpha})&=& 
-Q_0Q_5^2-3Q_0Q_5^3-5Q_0Q_5^4-3Q_0^2Q_5^3-7Q_0Q_5^5-25Q_0^2Q_5^4-112Q_0^2Q_5^5-5Q_0^3Q_5^4+\cdots~,\nonumber\\
A^{(1,3)}(Q_{\alpha})&=& Q_0Q_5^2+2Q_0Q_5^3+4Q_0Q_5^4+2Q_0^2Q_5^3+6Q_0Q_5^5+20Q_0^2Q_5^4+90Q_0^2Q_5^5+4Q_0^3Q_5^4+\cdots~,\nonumber\\
A^{(2,2)}(Q_{\alpha})&=& Q_0Q_5^2+2Q_0Q_5^3+4Q_0Q_5^4+2Q_0^2Q_5^3+6Q_0Q_5^5+17Q_0^2Q_5^4+78Q_0^2Q_5^5+4Q_0^3Q_5^4+\cdots~.\nonumber
\end{eqnarray}
These amplitudes agree with the topological vertex calculus with the framing ${\widetilde f}=-1$ in appendix A as far as we have checked.

\subsection{The annulus amplitude on ${\IC}^3/{{\IZ}_4}$ orbifold}%sec4.4

Here we consider the orbifold phase ${\IC}^3/{{\IZ}_4}$ of $K_{{\IF}_2}$ investigated in \cite{BrTan,Coat,ALM}, and predict the open orbifold Gromov-Witten invariants of ${\IC}^3/{{\IZ}_4}$.\footnote{In \cite{BrTan}(version 4), by using the remodeling approach summarized in section 2 and discussing modularity of the topological string amplitudes, the genus two closed string amplitude on ${\IC}^3/{{\IZ}_4}$ was computed. In this computation, the open obifold amplitudes $F_o^{(0,2)}$, $F_o^{(1,1)}$, $F_o^{(0,3)}$, $F_o^{(1,2)}$ and $F_o^{(2,1)}$ are also computed, and I was informed from Andrea Brini that these results will appear in the near future in \cite{BrCav} with the localization calculus by Renzo Cavalieri. My computation of $F_o^{(0,2)}=A_o$ completely agrees with the parallel computation by A. Brini.} Following \cite{BrTan,ALM}, let us consider a transformation $z_0=a_3^{-2},~z=a_1^{-2}$ of the closed string moduli. In these new coordinates $a_1$ and $a_3$, by considering the PF operators (\ref{eqn:3.17}),
$$
{\cal D}_0=\theta_{a_3}\left(\theta_{a_3}-1\right)-\frac{1}{16}a_3^2\left(\theta_{a_1}+2\theta_{a_3}\right)^2~,\quad{\cal D}_5=a_3\theta_{a_1}\left(\theta_{a_1}-1\right)+\frac12 a_1^2\theta_{a_1}\theta_{a_3}~,\quad \theta_{a_i}:=a_i\frac{\partial}{\partial a_i}~,
$$
we obtain the orbifold mirror maps\footnote{We can define a weight such that $a_1~(a_3)$ has weight $1/4~(1/2)$ with respect to the ${\IZ}_4$ (${\IZ}_2$) action with weights $(1,1,2)$. Then the following mirror map $s_{1/4}~(s_{1/2})$ also has weight $1/4~(1/2)$. Note that $s_{1/4}$ and $s_{1/2}$ respectively correspond to the blow up mode of the crepant partial resolution of ${\IC}^3/{\IZ}_4$ and the crepant resolution of the canonical bundle $K_{{\IP}(1,1,2)}$ \cite{Coat} as discussed for ${\IC}^3/{\IZ}_3$ in \cite{ABK,BKMP1,BKMP2}.}
\begin{eqnarray}
\label{eqn:4.53}%ラベル指定
s_{1/2}(a_3)&=& \sum_{k=0}^{\infty}\frac{\left[(2k-1)!!\right]^2}{4^k(2k+1)!}a_3^{2k+1}~,\\
s_{1/4}(a_{\alpha})&=& \sum_{m,k=0}^{\infty}\left[\frac{(4m-1)!!\Gamma(k+m+\frac14)^2}{2^{2m}(4m+1)!(2k)!\Gamma(\frac14)^2}a_1^{4m+1}a_3^{2k}-\frac{(4m+1)!!\Gamma(k+m+\frac54)^2}{2^{2m+1}(4m+3)!(2k+1)!\Gamma(\frac14)^2}a_1^{4m+3}a_3^{2k+1}\right]~.\nonumber
\end{eqnarray}
From (\ref{eqn:4.50}), we can obtain $G(\tau)$ in this orbifold phase replacing $T_z(z_{\alpha})$ with $s_{1/4}(a_{\alpha})$,
\begin{eqnarray}
&&\Delta_0(z_{\alpha})=-64a_1^{-4}\left(1-\frac{1}{64}(a_1^2-4a_3)^2\right)=:-64a_1^{-4}{\widetilde \Delta_0(a_{\alpha})}~,\hspace{15em}\nonumber\\
&&G(\tau)=-\frac{8}{3}a_1^{-4}{\widetilde \Delta_0(a_{\alpha})}\theta_{a_1} \left\{12\log \theta_{a_1} s_{1/4}(a_{\alpha})-12\log a_1 +\log {\widetilde \Delta_0(a_{\alpha})}\right\}~,\quad \theta_{a_1}:=a_1\frac{\partial}{\partial a_1}~.
\label{eqn:4.54}%ラベル指定
\end{eqnarray}
As in example 1-2, the open orbifold mirror map is given by 
\begin{equation}
X=-a_1(s_{1/2},s_{1/4})^{-1}x~,
\label{eqn:4.55}%ラベル指定
\end{equation}
with weight $-1/4$. Here we can compute the orbifold annulus amplitude $A_o(X_1,X_2)$ on ${\IC}^3/{{\IZ}_4}$,
\begin{equation}
A_o(X_1,X_2)=\int^{x_1}\int^{x_2} W_o^{(0,2)}(x_1^{-1},x_2^{-1})\frac{1}{x_1^2x_2^2} = \sum_{i_1,i_2}A_o^{(i_1,i_2)}(s_{1/2},s_{1/4})X_1^{i_1}X_2^{i_2}~,
\label{eqn:4.56}%ラベル指定
\end{equation}
where because $A_o(X_1,X_2)$ is the ${\IZ}_2$ and ${\IZ}_4$ monodromy invariant quantity, $A_o^{(i_1,i_2)}(s_{\alpha})=A_o^{(i_1,i_2)}(s_{1/2},s_{1/4})$ must have weight $(i_1+i_2)/4$. Concretely we obtain \vspace{-0.1em}
\begin{eqnarray}
&&A_o^{(1,1)}(s_{\alpha})= \left(-\frac14 s_{1/2}+\frac{1}{32}\frac{s_{1/2}^3}{3!}-\frac{1}{64}\frac{s_{1/2}^5}{5!}+\cdots\right)+\left(\frac{1}{8}+\frac{3}{256}\frac{s_{1/2}^4}{4!}+\frac{5}{128}\frac{s_{1/2}^6}{6!}+\cdots\right)\frac{s_{1/4}^2}{2!} \nonumber\\
&&\quad +\left(-\frac{1}{128}s_{1/2}-\frac{3}{128}\frac{s_{1/2}^3}{3!}-\frac{201}{2048}\frac{s_{1/2}^5}{5!}+\cdots\right)\frac{s_{1/4}^4}{4!}+\left(\frac{3}{128}+\frac{35}{512}\frac{s_{1/2}^2}{2!}+\frac{21}{64}\frac{s_{1/2}^4}{4!}+\cdots\right)\frac{s_{1/4}^6}{6!}+\cdots~,\nonumber\\
&&A_o^{(1,2)}(s_{\alpha})= \left(\frac18 s_{1/2}-\frac{5}{128}\frac{s_{1/2}^3}{3!}+\frac{21}{2048}\frac{s_{1/2}^5}{5!}+\cdots\right)s_{1/4}+\left(-\frac{3}{16}+\frac{5}{256}\frac{s_{1/2}^2}{2!}-\frac{35}{4096}\frac{s_{1/2}^4}{4!}+\cdots\right)\frac{s_{1/4}^3}{3!} \nonumber\\
&&\quad +\left(\frac{-3}{512}s_{1/2}+\frac{255}{8192}\frac{s_{1/2}^3}{3!}+\frac{15737}{131072}\frac{s_{1/2}^5}{5!}+\cdots\right)\frac{s_{1/4}^5}{5!}-\left(\frac{21}{1024}+\frac{2505}{16384}\frac{s_{1/2}^2}{2!}+\frac{189045}{262144}\frac{s_{1/2}^4}{4!}+\cdots\right)\frac{s_{1/4}^7}{7!}+\cdots~,\nonumber\\
&&A_o^{(1,3)}(s_{\alpha})= \left(\frac13 +\frac{1}{6}\frac{s_{1/2}^2}{2!}-\frac{1}{8}\frac{s_{1/2}^4}{4!}+\cdots\right)+\left(-\frac{5}{24}s_{1/2}+\frac{3}{32}\frac{s_{1/2}^3}{3!}-\frac{15}{256}\frac{s_{1/2}^5}{5!}+\cdots\right)\frac{s_{1/4}^2}{2!} \nonumber\\
&&\quad +\left(\frac{1}{2}-\frac{19}{192}\frac{s_{1/2}^2}{2!}+\frac{7}{128}\frac{s_{1/2}^4}{4!}+\cdots\right)\frac{s_{1/4}^4}{4!}+\cdots~,\nonumber\\
&&A_o^{(2,2)}(s_{\alpha})= \frac12+\left(-\frac{1}{8}s_{1/2} +\frac{1}{16}\frac{s_{1/2}^3}{3!}-\frac{7}{256}\frac{s_{1/2}^5}{5!}+\cdots\right)\frac{s_{1/4}^2}{2!}+\left(\frac{3}{8}-\frac{5}{64}\frac{s_{1/2}^2}{2!}+\frac{5}{256}\frac{s_{1/2}^4}{4!}+\cdots\right)\frac{s_{1/4}^4}{4!}+\cdots~.\nonumber
\end{eqnarray}
These amplitudes give physical predictions of the open orbifold Gromov-Witten invariants of ${\IC}^3/{\IZ}_4$.

\section{Higher amplitudes}%sec5

Here let us consider the higher amplitudes $F^{(1,1)}$ and $F^{(0,3)}$. From (\ref{eqn:2.3}), $W^{(1,1)}(p)$ is given by
\begin{equation}
W^{(1,1)}(p)=\sum_{q_i}\mathop{Res}_{q=q_i}\frac{dE_{q,\bar{q}}(p)}{\omega(q)-\omega(\bar{q})}B(q,\bar{q})~,
\label{eqn:4.57}%ラベル指定
\end{equation}
and therefore we must estimate $dE_{q,\bar{q}}(p)$, $\omega(q)$ and $B(q,\bar{q})$ around the ramification points $q=q_i$ or $s=s_i$ by a projected coordinate $s(q)=s({\bar q})$. Let us expand these quantities around $t^2=s-s_i=0$ as follows ;
\begin{eqnarray}
B(q,\bar{q})&=& -\frac14\left(\frac{G(\tau)}{\sigma(s)}-\frac{\sigma ''(s)}{3\sigma(s)}+\frac{\sigma '(s)^2}{4\sigma(s)^2}\right)dsds \hspace{22em}\nonumber\\
\label{eqn:4.58}%ラベル指定
& \simeq& -\left\{\frac{1}{4t^2}+\left(\frac{G(\tau)}{\sigma '(s_i)}-\frac{\sigma ''(s_i)}{12\sigma '(s_i)}\right)+{\cal O}(t^2)\right\}dtdt~,\\
dE_{q,\bar{q}}(p)& \simeq& -\frac{dx}{\sqrt{\sigma(x)}}\frac{t}{\sqrt{\sigma '(s_i)}}\left\{\left(\frac12 G(\tau)+\frac{f(x,s_i)}{(x-s_i)^2}\right)\right. \nonumber\\
\label{eqn:4.59}%ラベル指定
&&\hspace{0.5em}\left.+\frac{t^2}{6}\left(-\frac{\sigma ''(s_i)}{4\sigma '(s_i)}G(\tau)-\frac{\sigma ''(s_i)f(x,s_i)}{2\sigma '(x_i)(x-s_i)^2}+\frac{2\partial_sf(x,s_i)}{(x-s_i)^2}+\frac{4f(x,s_i)}{(x-s_i)^3}\right)+{\cal O}(t^4)\right\}~,\\
\label{eqn:4.60}%ラベル指定
\omega(q)-\omega(\bar{q})& \simeq& \frac{4t^2\sqrt{\sigma '(s_i)}}{\left(t^2+s_i\right)f(s_i)}dt \left\{1+t^2\left(\frac{\sigma ''(s_i)}{4\sigma '(s_i)}-\frac{f'(s_i)}{f(s_i)}+\frac{\sigma '(s_i)}{3f(s_i)^2}\right)+{\cal O}(t^4)\right\}~.
\end{eqnarray}
By these expansion we obtain the one-holed torus amplitude by the projected coordinate $x(p)=x({\bar p})$,
\begin{eqnarray}
\label{eqn:4.61}%ラベル指定
&&W^{(1,1)}(x)= \frac{dx}{8\sqrt{\sigma(x)}}\sum_{s_i}\left\{\frac{h(s_i)}{\sigma '(s_i)^2}G(\tau)^2+\left(\frac{2h(s_i)f(x,s_i)}{(x-s_i)^2\sigma '(s_i)^2}-\frac{h(s_i)\sigma ''(s_i)}{6\sigma '(s_i)^2}-\frac{s_i}{12f(s_i)}\right)G(\tau) \right.\hspace{5em}\\
&&\quad \left. +\left(\frac{h(s_i)f(x,s_i)}{3(x-s_i)^3\sigma '(s_i)}+\frac{3h'(s_i)f(x,s_i)+h(s_i)\partial_s f(x,s_i)}{6(x-s_i)^2\sigma '(s_i)}-\frac{h(s_i)\sigma ''(s_i)f(x,s_i)}{3(x-s_i)^2\sigma '(s_i)^2}-\frac{s_if(x,s_i)}{6(x-s_i)^2f(s_i)}\right)\right\}~,\nonumber
\end{eqnarray}
where we defined $h(x)=xf(x)$. By the same way, we can obtain $W^{(0,3)}(x_1,x_2,x_3)$ as follows,
\begin{eqnarray}
W^{(0,3)}(x_1,x_2,x_3)&=&2\sum_{q_i}\mathop{Res}_{q=q_i}\frac{dE_{q,\bar{q}}(p_1)}{\omega(q)-\omega(\bar{q})}B(p_2,q)B(p_3,\bar{q})\hspace{20em} \nonumber\\
&=& \frac{dx_1dx_2dx_3}{16\sqrt{\sigma(x_1)\sigma(x_2)\sigma(x_3)}}\sum_{s_i}\frac{h(s_i)}{\sigma '(s_i)^2}\left\{G(\tau)^3+2\left(\frac{f(x_1,s_i)}{(x_1-s_i)^2}+\mbox{cyclic}\right)G(\tau)^2\right. \nonumber\\
&&\left.+4\left(\frac{f(x_1,s_i)f(x_2,s_i)}{(x_1-s_i)^2(x_2-s_i)^2}+\mbox{cyclic}\right)G(\tau)+\frac{8f(x_1,s_i)f(x_2,s_i)f(x_3,s_i)}{(x_1-s_i)^2(x_2-s_i)^2(x_3-s_i)^2}\right\}~.
\label{eqn:4.62}%ラベル指定
\end{eqnarray}
In appendix D we compute the one-holed torus amplitudes $F^{(1,1)}$ on $K_{{\IF}_0}$ and $K_{{\IF}_2}$ on the mirror side, and we see (\ref{eqn:2.16}) explicitly. In appendix E we compute the one-holed torus amplitudes $F^{(1,1)}$ and genus zero, three-hole amplitudes $F^{(0,3)}$ on $K_{{\IF}_2}$ and its orbifold phase ${\IC}^3/{\IZ}_4$.

\section{Conclusion}%sec6

In this paper we have studied the topological strings on the large class of local toric del Pezzo surfaces described in section 4 by the remodeling approach of \cite{BKMP1}, and as the extension of the result proved in the case of $K_{{\IP}^2}$ in \cite{BKMP2}, we proved the functional formula (\ref{eqn:4.6}) written by the period to obtain the annulus amplitudes on these geometries. Based on this formula, we computed the annulus amplitudes on $K_{{\IF}_0}$, $K_{dP_2^{(0)}}$, $K_{{\IF}_2}$ and their orbifold phase $T^*S^3/{\IZ}_2$, ${\IC}^3/{\IZ}_4$. The annulus amplitude on ${\IC}^3/{\IZ}_4$ gives physical predictions of the open orbifold Gromov-Witten invariants, and further computation will be summarized in appendix E. It is interesting to discuss generalizations of the formula (\ref{eqn:2.13}) and (\ref{eqn:4.6}) as follows ;

$\bullet$ This formula is not applicable for the local toric surfaces described by the dashed lines in figure \ref{fig:1}, because these mirror curves cannot be written in the form (\ref{eqn:2.6}). We want to know more universal formula for genus one mirror curves, so we should discuss how to overcome this problem.

$\bullet$ When we consider ladder diagrams constructed from local ${\IF}_m$ as \cite{IqKP2}, the genus of these mirror curves is larger than one, so we need the Bergman kernel for spectral curves with ${\bar g}\ge 2$. We do not know this explicit form, but it is interesting to consider a generalization of (\ref{eqn:2.13}) and (\ref{eqn:4.6}) for these geometries.

These generalizations should be useful to understand the BKMP conjecture (\ref{eqn:3.13}) and the structure of the local B-model geometry with open string moduli.

Via geometric engineering \cite{KKVa}, the local toric surfaces in figure \ref{fig:1} are related to the Seiberg-Witten geometries of supersymmetric $SU(2)$ gauge theory on ${\IR}^4\times S^1$ with or without several fundamental hypermultiplets. It is also interesting to discuss relations between phase (or moduli) spaces of these geometries, i.e. the local B-model geometry \cite{HKR,ALM} versus the Seiberg-Witten geometry \cite{HuKl0,HuKl}.\\

\noindent{\large {\bf Acknowledgements:}} I would like to thank Hidetoshi Awata, Hiroyuki Fuji, Satoshi Minabe, Sanefumi Moriyama and Naoto Yotsutani for useful discussions and comments. Especially I would also like to thank Hiroaki Kanno for enlightening discussions, comments and careful reading of the manuscript. I would also grateful to Andrea Brini for comments on the previous version of this paper and sharing his unpublished results of the topological open string amplitudes on ${\IC}^3/{\IZ}_4$ orbifold \cite{BrCav}.

\appendix

\section{Topological vertex calculus (or the calculus on the A-side)}%secA

In this appendix, we summarize how to compute ${\cal F}_A^{(g,h)}$ of (\ref{eqn:3.5}) by the topological vertex \cite{AgKMVa,Mar2}. Based on the fact that there is the one to one correspondence between the representation space of $U(\infty)$ and the representation space of symmetric group $\sum_{d=1}^{\infty}S_d$ or the Young diagram, let us rewrite (\ref{eqn:3.5}) by the Frobenius's character formula
\begin{equation}
Tr_R V=\sum_{\vec k}\frac{1}{z_{\vec k}}\chi_R(C({\vec k}))\Upsilon_{\vec k}(V)~,\quad z_{\vec k}:=\prod_jk_j!j^{k_j}~,\quad \Upsilon_{\vec k}(V):=\prod\left(TrV^j\right)^{k_j}~,
\label{eqn:A.1}%ラベル指定
\end{equation}
where ${\vec k}=(k_1,k_2,\cdots)$ which $k_i$ is the number of $i$ included in ${\vec \omega}$ is the winding base, and $\chi_R(C({\vec k}))$ is the character with respect to the conjugation class $C({\vec k})$. Via the winding base, (\ref{eqn:3.5}) is rewritten as
\begin{eqnarray}
\label{eqn:A.2}%ラベル指定
{\cal F}(V)&=&\sum_{g=0}^{\infty}\sum_{\vec k}\frac{1}{\prod k_j!}g_s^{2g-2+h}{\cal F}_{{\vec k},g}(Q)\Upsilon_{\vec k}(V) \\
&=&\sum_R{\cal F}_R(g_s,Q)Tr_R V=\log\Bigl\{1+\frac{1}{Z(g_s,Q)}\sum_{R \neq \bullet}Z_R(g_s,Q)Tr_R V \Bigr\}~,
\label{eqn:A.3}%ラベル指定
\end{eqnarray}
where $\bullet$ represents the trivial representation of $U(\infty)$ and $Z(g_s,Q)=Z_{\bullet}(g_s,Q)$. Here $Z_R(g_s,Q)$ can be computed by the topological vertex, and therefore by (\ref{eqn:A.1}), we can compute ${\cal F}_A^{(g,h)}$. For example, by the box expansion of (\ref{eqn:A.3}) we obtain
\begin{eqnarray}
&&{\cal F}_A^h(X_i;Q):=\sum_{g=0}^{\infty}g_s^{2g-2+h}{\cal F}_A^{(g,h)}(X_1,\cdots,X_h;Q)~,\nonumber\\
&&{\cal F}_A^1(X_1;Q)=\frac{1}{Z}Z_{\tableau{1}}X_1+\frac{1}{2Z}\Bigl[Z_{\tableau{2}}-Z_{\tableau{1 1}}\Bigl]X_1^2+\frac{1}{3Z}\Bigl[Z_{\tableau{3}}-Z_{\tableau{2 1}}+Z_{\tableau{1 1 1}}\Bigl]X_1^3 \nonumber\\
&&\hspace{6em} +\frac{1}{4Z}\Bigl[Z_{\tableau{4}}-Z_{\tableau{3 1}}+Z_{\tableau{2 1 1}}-Z_{\tableau{1 1 1 1}}\Bigl]X_1^4+\cdots~,\nonumber\\
&&{\cal F}_A^2(X_i;Q)=\Bigl[\frac{1}{Z}(Z_{\tableau{2}}+Z_{\tableau{1 1}})-\frac{1}{Z^2}Z_{\tableau{1}}^2\Bigl]X_1X_2+\frac12\Bigl[\frac{1}{Z}(Z_{\tableau{3}}-Z_{\tableau{1 1 1}})-\frac{1}{Z^2}Z_{\tableau{1}}(Z_{\tableau{2}}-Z_{\tableau{1 1}})\Bigl](X_1X_2^2+X_1^2X_2) \nonumber\\
&&\hspace{6em} +\frac13\Bigl[\frac{1}{Z}(Z_{\tableau{4}}-Z_{\tableau{2 2}}+Z_{\tableau{1 1 1 1}})-\frac{1}{Z^2}Z_{\tableau{1}}(Z_{\tableau{3}}-Z_{\tableau{2 1}}+Z_{\tableau{1 1 1}})\Bigl](X_1X_2^3+X_1^3X_2) \nonumber\\
&&\hspace{6em} +\frac14\Bigl[\frac{1}{Z}(Z_{\tableau{4}}-Z_{\tableau{3 1}}+2Z_{\tableau{2 2}}-Z_{\tableau{2 1 1}}+Z_{\tableau{1 1 1 1}})-\frac{1}{Z^2}(Z_{\tableau{2}}-Z_{\tableau{1 1}})^2\Bigl]X_1^2X_2^2+\cdots~,\nonumber\\
&&{\cal F}_A^3(X_i;Q)=\Bigl[\frac{1}{Z}(Z_{\tableau{3}}+2Z_{\tableau{2 1}}+Z_{\tableau{1 1 1}})-\frac{3}{Z^2}Z_{\tableau{1}}(Z_{\tableau{2}}+Z_{\tableau{1 1}})+\frac{2}{Z^3}Z_{\tableau{1}}^3\Bigl]X_1X_2X_3 \nonumber\\
&&\hspace{6em} +\frac12\Bigl[\frac{1}{Z}(Z_{\tableau{4}}+Z_{\tableau{3 1}}-Z_{\tableau{2 1 1}}-Z_{\tableau{1 1 1 1}})-\frac{1}{Z^2}(Z_{\tableau{2}}^2-Z_{\tableau{1 1}}^2+2Z_{\tableau{1}}(Z_{\tableau{3}}-Z_{\tableau{1 1 1}})) \nonumber\\
&&\hspace{21.5em} +\frac{2}{Z^3}Z_{\tableau{1}}^2(Z_{\tableau{2}}-Z_{\tableau{1 1}})\Bigl](X_1X_2X_3^2+\mbox{cyclic})+\cdots~.\nonumber
\end{eqnarray}
In this way, we can compute the A-model amplitudes ${\cal F}_A^{(g,h)}$ for an arbitrary local toric Calabi-Yau threefold with an A-brane. If we want to consider the framing ambiguity ${\widetilde f} \in {\IZ}$, by\footnote{$\mu_R$ is the Young diagram corresponding to the representation $R$.}
\begin{equation}
Z_{R}(g_s,Q)~ \longrightarrow ~ q^{{\widetilde f} \kappa_R/2}Z_{R}(g_s,Q)~,\quad q=e^{g_s}~,~\kappa_R=2\sum_{(i,j)\in \mu_R}(j-i)~,
\label{eqn:A.4}%ラベル指定
\end{equation}
we obtain the A-model amplitudes with the framing ${\widetilde f}$.

\section{Torus amplitudes on local toric del Pezzo surfaces}%secB

In this appendix, we discuss the torus ($g=1,h=0$) amplitudes \cite{Mar1} for 11 local toric surfaces considered in section 4. In the context of the Hermitian one-matrix model, Akemann explicitly wrote down the torus amplitude $F^{(1,0)}$ by solving the loop equation under two-cut ansatz \cite{Ake} (see \cite{Chek} for multi-cut solution). For the mirror curve (\ref{eqn:4.35}) of $K_{dP_5}$,
\begin{equation}
\sigma(x)=\frac{1}{1-4z_0}\left\{(x^2-x+z_0^{-\frac12}z)^2-4\biggl(z_0x^4+\sum_{\alpha=1}^4 z^{\frac{\alpha}{2}}z_0^{1-\frac{\alpha}{2}}z_{\alpha}(-x)^{4-\alpha}\biggl)\right\}=\prod_{i=1}^4(x-s_i)~,
\label{eqn:B.1}%ラベル指定
\end{equation}
let us use the formula
\begin{eqnarray}
\label{eqn:B.2}%ラベル指定
F^{(1,0)}&=&-\frac12 \log \omega_1-\frac{1}{24}\log|M_1M_2M_3M_4|-\frac{1}{12}\log |\Delta(z_{\alpha})|~,\quad \omega_1=\frac{2i}{\pi}\frac{K(k)}{\sqrt{(s_1-s_3)(s_2-s_4)}}~,\hspace{2.5em}\\
M(x)&=&\frac{1}{x \sqrt{\sigma(x)}}\tanh^{-1} \left[\frac{(1-4z_0)^{\frac12}\sqrt{\sigma(x)}}{x^2-x+z_0^{-\frac12}z}\right]~,\quad M_i=\frac{(1-4z_0)^{\frac12}}{s_i(s_i^2-s_i+z_0^{-\frac12}z)}~,
\label{eqn:B.3}%ラベル指定
\end{eqnarray}
where $\Delta(z_{\alpha})=\prod_{i<j}(s_i-s_j)^2$ is the discriminant of the curve (\ref{eqn:B.1}), and $M_i=M(s_i)$ are called the first moments in the context of the matrix model. From (\ref{eqn:4.34}), we get
\begin{equation}
\omega_1=i(1-4z_0)^{\frac12}\theta T_z(z_{\alpha})~,
\label{eqn:B.4}%ラベル指定
\end{equation}
therefore we obtain
\begin{equation}
F^{(1,0)}=-\frac12\log \theta T_z(z_{\alpha})-\frac{1}{24}\log \frac{z_0(1-4z_0)^9\Delta(z_{\alpha})^2}{z^2(1-4z_4)}\prod_{i=1}^4\frac{1}{s_i^2-s_i+z_0^{-\frac12}z}=:-\frac12\log \theta T_z(z_{\alpha})-\frac{1}{24}\log f^{(1,0)}~,
\label{eqn:B.5}%ラベル指定
\end{equation}
where $f^{(1,0)}$ is corresponding to a holomorphic ambiguity of the BCOV holomorphic anomaly equation \cite{BCOV1,BCOV2}. Though solutions to the BCOV holomorphic anomaly equation are non-holomorphic quantities, but by taking the holomorphic limit, we can obtain holomorphic quantities, especially as the torus amplitude \cite{KlZas},
\begin{equation}
{\cal F}_B^{(1,0)}=-\frac12 \log \det \left(\theta_{\alpha}T_{\beta}\right)+\log \Delta_{discrim}~,\quad \theta_{\alpha}:=z_{\alpha}\frac{\partial}{\partial z_{\alpha}}~,
\label{eqn:B.6}%ラベル指定
\end{equation}
in the case of local toric Calabi-Yau threefold. Where $\Delta_{discrim}$ is the holomorphic ambiguity given by a combination of the discriminant of the characteristic variety obtained from the PF equations (\ref{eqn:3.17}). In order to obtain solutions to the BCOV holomorphic anomaly equation, we must fix the holomorphic ambiguity such as $\Delta_{discrim}$ by considering appropriate boundary conditions which are called the gap conditions \cite{GhV,HKQ}. Whereas the first term of (\ref{eqn:B.6}) is computed via the PF equations (\ref{eqn:3.17}). In the case of $K_{dP_5}$, from (\ref{eqn:4.36}) we obtain
\begin{eqnarray}
Jac(z_{\alpha}):&=& \det \left(\theta_{\alpha}T_{\beta}\right)=\sum_{\sigma\in S_6}sgn(\sigma)\theta_0T_{\sigma(0)}\theta_1T_{\sigma(1)}\theta_2T_{\sigma(2)}\theta_3T_{\sigma(3)}\theta_4T_{\sigma(4)}\theta_5T_{\sigma(5)} \nonumber\\
&=& \left(1-4z_0\right)^{-\frac12} \theta T_z(z_{\alpha})~.
\label{eqn:B.7}%ラベル指定
\end{eqnarray}
We can see that $\omega_1$ of (\ref{eqn:B.2}) is surely corresponding to $Jac(z_{\alpha})$. In the rest of this appendix we compute the genus one A-model amplitudes ${\cal F}_A^{(1,0)}$ on $K_{{\IF}_0},~K_{{\IF}_1}$ and $K_{{\IF}_2}$ (corresponding to no.2, no.3 and no.4 of figure \ref{fig:1} in section 1) based on (\ref{eqn:B.5}) and the BKMP conjecture (\ref{eqn:3.13}). A mirror curve of these varieties and the discriminant are given by
\begin{eqnarray}
\label{eqn:B.8}%ラベル指定
K_{{\IF}_0}:\quad\quad && y^2-(x^2-x+z)y+zz_2x^2=0~,\quad \sigma(x)=(x^2-x+z)^2-4zz_2x^2~,\hspace{8em}\\
&&\Delta(z_{\alpha})=256z^4z_2^2\Delta_0(z_{\alpha})~,\quad \Delta_0(z_{\alpha}):=1-8z(1+z_2)+16z^2(1-z_2)^2~,\hspace{8em}\nonumber\\
\label{eqn:B.9}%ラベル指定
K_{{\IF}_1}:\quad\quad && y^2-(x^2-x+z)y-z^{\frac12}z_1x^3=0~,\quad \sigma(x)=(x^2-x+z)^2+4z^{\frac12}z_1x^3~,\\
&&\Delta(z_{\alpha})=256z^4z_1^2\Delta_0(z_{\alpha})~,\quad \Delta_0(z_{\alpha}):=(1-4z)^2-z^{\frac12}z_1(1-36z+27z^{\frac32}z_1)~,\nonumber\\
\label{eqn:B.10}%ラベル指定
K_{{\IF}_2}:\quad\quad && y^2-(x^2-x+z_0^{-\frac12}z)y+z_0x^4=0~,\quad \sigma(x)=\frac{1}{1-4z_0}\left\{(x^2-x+z_0^{-\frac12}z)^2-4z_0x^4\right\}~,\\
&&\Delta(z_{\alpha})=\frac{256z^4\Delta_0(z_{\alpha})}{(1-4z_0)^6}~,\quad \Delta_0(z_{\alpha}):=(1-4z_0^{-\frac12}z)^2-64z^2~.\nonumber
\end{eqnarray}
Here we can easily compute $f^{(1,0)}$ of (\ref{eqn:B.5}) as
\begin{eqnarray}
\label{eqn:B.11}%ラベル指定
K_{{\IF}_0}&:&\quad\quad f^{(1,0)}=4^6z^2z_2^2\Delta_0(z_{\alpha})^2~,\\
\label{eqn:B.12}%ラベル指定
K_{{\IF}_1}&:&\quad\quad f^{(1,0)}=4^6z^2z_1^2\Delta_0(z_{\alpha})^2~,\\
\label{eqn:B.13}%ラベル指定
K_{{\IF}_2}&:&\quad\quad f^{(1,0)}=\frac{4^6z_0z^2\Delta_0(z_{\alpha})^2}{1-4z_0}~,
\end{eqnarray}
and from (\ref{eqn:4.36}) the mirror maps are given by
\begin{eqnarray}
\label{eqn:B.14}%ラベル指定
K_{{\IF}_0}&:&\quad T_z(z_{\alpha})=\log z+2\sum_{m,n\ge 0}^{\infty}\frac{(2m+2n-1)!}{m!^2n!^2}z^{m+n}z_2^n~,\quad T_2(z_2)=\log z_2~,\hspace{12em}\\
\label{eqn:B.15}%ラベル指定
K_{{\IF}_1}&:&\quad T_z(z_{\alpha})=\log z+2\sum_{m,n \ge 0,~m \ge n}^{\infty}\frac{(-1)^n(2m+n-1)!}{m!n!^2(m-n)!}z^{m+\frac{n}{2}}z_1^n~,\quad T_1(z_1)=\log z_1~,\\
\label{eqn:B.16}%ラベル指定
K_{{\IF}_2}&:&\quad T_z(z_{\alpha})=\log z+2\sum_{m,n \ge 0,~m \ge 2n}^{\infty}\frac{(2m-1)!}{m!n!^2(m-2n)!}z^mz_0^{-\frac{m}{2}+n}~,\quad T_0(z_0)=\log \frac{1-\sqrt{1-4z_0}}{1+\sqrt{1-4z_0}}~.
\end{eqnarray}
From these we obtain the genus one A-model amplitude ${\cal F}_A^{(1,0)}$ from (\ref{eqn:B.5}), and by a change of variables of (\ref{eqn:4.30}) as $Q_z\to Q_5$, $Q_2\to Q_2/Q_5$ for $K_{{\IF}_0}$, $Q_z\to Q_5$, $Q_1 \to Q_1/Q_5^{1/2}$ for $K_{{\IF}_1}$ and $Q_z \to Q_0^{1/2}Q_5$, $Q_0\to Q_0$ for $K_{{\IF}_2}$, we obtain ${\cal F}_A^{(1,0)}$ in the original variables as follows ;
\begin{eqnarray}
K_{{\IF}_0}: {\cal F}_A^{(1,0)}&=&-\frac12\log \theta T_z(z_{\alpha})-\frac{1}{12}\log zz_2\Delta_0(z_{\alpha}) \nonumber\\
&=&-\frac{1}{12}\log Q_2-\left(\frac16 Q_5+\frac{1}{12} Q_5^2+\frac{1}{18}Q_5^3+\cdots\right)-\left(\frac16+\frac13 Q_5 +\frac12 Q_5^2 +\frac23 Q_5^3+\cdots\right)Q_2 \nonumber\\
&&\hspace{-1em}-\left(\frac{1}{12}+\frac12 Q_5 -\frac{37}{6} Q_5^2 -\frac{353}{6} Q_5^3+\cdots\right)Q_2^2-\left(\frac{1}{18}+\frac23 Q_5 -\frac{353}{6} Q_5^2 -\frac{8576}{9} Q_5^3+\cdots\right)Q_2^3+\cdots \nonumber\\
\label{eqn:B.17}%ラベル指定
&=&-\frac{1}{12}\log Q_2+\frac16\log (1-Q_5)+\frac16\sum_{n=1}^{\infty}\frac{Q_2^n}{n(1-Q_5)^{4n}}H_n^{(0)}(Q_5)~,\\
K_{{\IF}_1}: {\cal F}_A^{(1,0)}&=&-\frac12\log \theta T_z(z_{\alpha})-\frac{1}{12}\log zz_1\Delta_0(z_{\alpha}) \nonumber\\
&=&-\frac{1}{12}\log Q_1Q_5^{\frac12}-\left(\frac16 Q_5+\frac{1}{12} Q_5^2+\frac{1}{18}Q_5^3+\cdots\right)+\left(\frac{1}{12}+\frac14 Q_5 +\frac{7}{12} Q_5^3 +\frac34 Q_5^4+\cdots\right)Q_1 \nonumber\\
&&\hspace{-1em}+\left(\frac{1}{24}-\frac38 Q_5^2 +\frac{19}{3} Q_5^3 +\frac{1417}{24} Q_5^4+\cdots\right)Q_1^2+\left(\frac{1}{36}-\frac{23}{3} Q_5^3 -\frac{1585}{6} Q_5^4+\cdots\right)Q_1^3+\cdots \nonumber\\
\label{eqn:B.18}%ラベル指定
&=&-\frac{1}{12}\log Q_1Q_5^{\frac12}+\frac16\log (1-Q_5)+\frac16\sum_{n=1}^{\infty}\frac{Q_1^n}{n(1-Q_5)^{4n}}H_n^{(1)}(Q_5)~,\\
K_{{\IF}_2}: {\cal F}_A^{(1,0)}&=&-\frac12\log \theta T_z(z_{\alpha})+\frac{1}{24}\log (1-4z_0)-\frac{1}{12}\log z_0^{\frac12}z\Delta_0(z_{\alpha}) \nonumber\\
&=&-\frac{1}{12}\log Q_0Q_5-\left(\frac16 Q_5+\frac{1}{12} Q_5^2+\frac{1}{18}Q_5^3+\cdots\right)-\left(\frac{1}{12}+\frac16 Q_5 +\frac13 Q_5^2+\frac12 Q_5^3+\cdots\right)Q_0 \nonumber\\
&&\hspace{-1em}-\left(\frac{1}{24}+\frac{1}{12} Q_5^2 +\frac12 Q_5^3 -\frac{37}{6}Q_5^4+\cdots\right)Q_0^2-\left(\frac{1}{36}+\frac{1}{18} Q_5^3 +\frac23 Q_5^4 -\frac{353}{6}Q_5^5+\cdots\right)Q_0^3+\cdots \nonumber\\
\label{eqn:B.19}%ラベル指定
&=&-\frac{1}{12}\log Q_0Q_5+\frac16\log (1-Q_5)+\frac16\sum_{n=1}^{\infty}\frac{Q_0^n}{n(1-Q_5)^{4n}}H_n^{(2)}(Q_5)~,
\end{eqnarray}
where $H_n^{(0)}(x),~H_n^{(1)}(x)$ and $H_n^{(2)}(x)$ are polynomials of degree $2n$, $3n$ and $4n$ respectively. These polynomials have the following symmetric forms,
\begin{eqnarray}
H_1^{(0)}(x)&=&1-2x+x^2~,\nonumber\\
H_2^{(0)}(x)&=&1-2x-94x^2-2x^3+x^4~,\nonumber\\
H_3^{(0)}(x)&=&1-1137x^2-3872x^3-1137x^4+x^6~,\nonumber\\
H_4^{(0)}(x)&=&1+4x-6818x^2-72168x^3-158262x^4+\mbox{symmetric}~,~\cdots ~,\nonumber\\
H_1^{(1)}(x)&=&\frac12-\frac12 x-\frac12 x^2+\frac12 x^3~,\nonumber\\
H_2^{(1)}(x)&=&\frac12-4x+\frac{19}{2}x^2+84x^3+\frac{19}{2}x^4-4x^5+\frac12 x^6~,\nonumber\\
H_3^{(1)}(x)&=&\frac12-6x+33x^2-248x^3-\frac{5703}{2}x^4-\frac{5703}{2}x^5-248x^6+33x^7-6x^8+\frac12 x^9~,\nonumber\\
H_4^{(1)}(x)&=&\frac12-8x+60x^2-280x^3+\frac{12131}{2}x^4+76868x^5+150812x^6+\mbox{symmetric}~,~\cdots~, \nonumber\\
H_1^{(2)}(x)&=&\frac12-x+x^2-x^3+\frac12 x^4~,\nonumber\\
H_2^{(2)}(x)&=&\frac12-4x+15x^2-30x^3-59x^4-30x^5+15x^6-4x^7+\frac12 x^8~,\nonumber\\
H_3^{(2)}(x)&=&\frac12-6x^2+33x^2-109x^3+\frac{495}{2}x^4-1533x^5-3410x^6+\mbox{symmetric}~,\nonumber\\
H_4^{(2)}(x)&=&\frac12-8x+60x^2-280x^3+911x^4-2180x^5-2814x^6-77888x^7-151827x^8+\mbox{symmetric}~,~\cdots~. \nonumber
\end{eqnarray}
These amplitudes completely agree with the topological vertex calculus as far as we have checked, and satisfy $|H_n^{(0)}(1)|=|H_n^{(1)}(1)|=|H_n^{(2)}(1)|$ expected from the geometric engineering limit \cite{KKVa,KMT}.

\section{Framing ambiguity}%secC

In this appendix, we give a relation of the open flat coordinate between an open string phase I and another open string phase {\II} \cite{BKMP1}. This is summarized in the following diagram.
\begin{center}
 \begin{tabular}{lccc}
 & Phase I & & Phase {\II}\\
A:\quad & $(X,Y(X))$ &\quad $\stackrel{f_A}{\longrightarrow}$\quad~ & $({\widetilde X},{\widetilde Y}({\widetilde X}))$\\
 & \rotatebox{90}{$\stackrel{\mbox{\rotatebox{270}{$g_1$}}}{\longleftarrow}$} & & \rotatebox{90}{$\stackrel{\mbox{\rotatebox{270}{$g_2$}}}{\dashleftarrow}$}\\
B:\quad & $(x,y(x))$ &\quad $\stackrel{f_B}{\longrightarrow}$\quad~ & $({\widetilde x},{\widetilde y}({\widetilde x}))$
 \end{tabular}
\end{center}
Where $f_A,~f_B$ denote the open string phase transition given by (\ref{eqn:4.5}), and $g_1,~g_2$ denote the open string mirror map given by (\ref{eqn:3.16}). Now as the phase I, we take a zero framing phase as considered in section 4, then we can write $g_1$ explicitly in the form
\begin{equation}
g_1^{-1}:\quad X=e^{\Delta_u}x~,\quad Y=y~.
\label{eqn:C.1}%ラベル指定
\end{equation}
Therefore from the above diagram, we can get the open mirror map $g_2$ in the phase {\II}. For example, let us consider a phase with the framing $f \in{\IZ}$ as the phase {\II}, 
\begin{eqnarray}
\label{eqn:C.2}%ラベル指定
f_A&:&\quad X_f:={\widetilde X}=XY^f~,\quad Y_f:={\widetilde Y}=Y~,\\
f_B&:&\quad x_f:={\widetilde x}=xy^f~,\quad y_f:={\widetilde y}=y~.
\label{eqn:C.3}%ラベル指定
\end{eqnarray}
As a result, we obtain
\begin{equation}
g_2^{-1}=f_A \circ g_1^{-1}\circ f_B^{-1}:\quad X_f=XY^f=e^{\Delta_u}xy^f=e^{\Delta_u}x_f~,\quad Y_f=Y=y=y_f~.
\label{eqn:C.4}%ラベル指定
\end{equation}
This has the same form as (\ref{eqn:C.1}). As an example, let us consider $K_{{\IF}_2}$ in subsection 4.3. By taking into account the inverse reparametrization $x \to x^{-1},~y \to x^{-2}y$ in the reparametrized mirror curve (\ref{eqn:4.48}), the open string modulus $(x_f,y_f(x_f))$ in this framing phase is obtained by
\begin{equation}
x=x_f^{-1}y_f^f~,\quad y=x_f^{-2}y_f^{2f+1}~,
\label{eqn:C.5}%ラベル指定
\end{equation}
and therefore (\ref{eqn:4.48}) is converted to
\begin{eqnarray}
\label{eqn:C.6}%ラベル指定
&&y_f^{4f+2}-y_f^{4f+1}+x_fy_f^{3f+1}-z_0^{-\frac12}zx_f^2y_f^{2f+1}+z^2x_f^4=0~,\hspace{10em}\\
&&y_f=1-x_f-(f-z_0^{-\frac12}z)x_f^2-\frac{f}{2}(3f-6z_0^{-\frac12}z+1)x_f^3+\cdots~.
\label{eqn:C.7}%ラベル指定
\end{eqnarray}
From (\ref{eqn:C.5}) we obtain $x=x(x_f)$, and because the Bergman kernel does not depend on a particular reparametrization of the mirror curve as noticed in (\ref{eqn:4.3}), the annulus amplitude with the framing $f$ is easily obtained from \vspace{-0.1em}
\begin{equation}
A_A(X_{f,1},X_{f,2})=\int^{x_{f,1}}\int^{x_{f,2}} B \left(x_1(x_{f,1}),x_2(x_{f,2})\right)-\frac{dx_{f,1}dx_{f,2}}{(x_{f,1}-x_{f,2})^2}~.
\label{eqn:C.8}%ラベル指定
\end{equation}
In appendix E we compute the disk and annulus amplitudes on $K_{{\IF}_2}$ and ${\IC}^3/{\IZ}_4$ with the framing $f$.

\section{Summary of the higher B-model amplitudes $F^{(1,1)}$}%secD

When we compute the higher amplitudes, we must consider a summation over ramification points as (\ref{eqn:4.61}) and (\ref{eqn:4.62}). This is carried out by the Cauchy's residue formula. As an example, we compute
\begin{equation}
\sum_{s_i}\frac{h(s_i)}{\sigma '(s_i)^2}~.
\label{eqn:D.1}%ラベル指定
\end{equation}
For this purpose, let us define ${\widetilde \sigma}(x;s_i)$ as
\begin{eqnarray}
\label{eqn:D.2}%ラベル指定
&&\sigma(x)=\prod_{i=1}^4(x-s_i)=(x-s_i)\prod_{j \ne i}(x-s_j)=:(x-s_i){\widetilde \sigma}(x;s_i)~,\\
&&{\widetilde \sigma}(x;s_i)=x^3-S_1(s_i)x^2+S_2(s_i)x-S_3(s_i)~,\nonumber\\
&&S_1(s_i)=S_1-s_i~,~S_2(s_i)=S_2-S_1(s_i)s_i~,~S_3(s_i)=S_3-S_2(s_i)s_i~.\nonumber
\end{eqnarray}
The discriminant $\Delta$ of $\sigma(x)$ and the discriminant $\Delta(s_i)$ of ${\widetilde \sigma}(x;s_i)$ are related as
\begin{equation}
\Delta=\prod_{i=1}^4\sigma'(s_i)=\sigma'(s_i)\prod_{j\ne i}\sigma'(s_j)=\sigma'(s_i)\prod_{j\ne i}(s_j-s_i){\widetilde \sigma}'(s_j;s_i)=\sigma'(s_i)^2\cdot (-1)\prod_{j\ne i}{\widetilde \sigma}'(s_j;s_i)=\sigma '(s_i)^2\Delta(s_i)~.
\label{eqn:D.3}%ラベル指定
\end{equation}
By this relation we can compute (\ref{eqn:D.1}) ;
$$
\sum_{s_i}\frac{h(s_i)}{\sigma '(s_i)^2}=\frac{1}{\Delta}\sum_{s_i}\frac{h(s_i)\Delta(s_i)\sigma '(s_i)}{\sigma '(s_i)}=\frac{1}{\Delta}\oint_{\cal C}\frac{dx}{2\pi i}\frac{h(x)\Delta(x)\sigma '(x)}{\sigma(x)} = \frac{1}{\Delta}\oint_0 \frac{dx}{2\pi i}\frac{h(x^{-1})\Delta(x^{-1})\sigma '(x^{-1})}{\sigma(x^{-1})}\frac{1}{x^2}~,
$$
where ${\cal C}$ denotes a contour around the branch points. In this way we can compute (\ref{eqn:4.61}) and (\ref{eqn:4.62}). For examples of subsection 4.3, we obtain the one-holed torus amplitude (\ref{eqn:4.61}) as follows ;\\
\noindent$\underline{\mbox{{\bf Example 1:}}~K_{{\IF}_0}}$\\
We use the mirror curve (\ref{eqn:4.39}). The one-holed torus amplitude is
\begin{eqnarray}
\label{eqn:D.4}%ラベル指定
W^{(1,1)}(x)&=&-\frac{32z^4z_2^2dx}{\sqrt{\sigma(x)}\Delta(z_{\alpha})}\left(2G(\tau)^2+f_1^{(1,1)}(x)G(\tau)+f_0^{(1,1)}(x)\right)~, \\
f_1^{(1,1)}(x)&=&-\frac{x(x^2-x+z)}{\sigma(x)}\Delta_0(z_{\alpha})~,\quad \Delta_0(z_{\alpha})=1-8z(1+z_2)+16z^2(1-z_2)^2~,\nonumber\\
f_0^{(1,1)}(x)&=&\frac{1}{9\sigma(x)^2}\left[\left\{1-8z(1+z_2)+16z^2(1-4z_2+z_2^2)\right\}x^8-4\left\{1-z(8+5z_2)\right.\right.\nonumber\\
&&\hspace{-1em}\left.+8z^2(2-11z_2-z_2^2)+48z^3z_2(1-z_2)^2\right\}x^7+2\left\{3-2z(11+17z_2)+32z^2(1-2z_2)^2\right. \nonumber\\
&&\hspace{-1em} \left.+32z^3(1-9z_2+15z_2^2-5z_2^3)\right\}x^6-4\left\{1-z(5+18z_2)-z^2(8+35z_2-120z_2^2)\right.\nonumber\\
&&\hspace{-1em}\left.+8z^3(6-17z_2+65z_2^2-44z_2^3)-48z^4z_2(9-8z_2)(1-z_2)^2\right\}x^5+\left\{1+4z(1-4z_2)\right. \nonumber\\
&&\hspace{-1em}\left. -2z^2(37+140z_2-48z_2^2)+16z^3(9+33z_2+128z_2^2-16z_2^3)\right. \nonumber\\
&&\hspace{-1em} \left.+32z^4(3-104z_2+211z_2^2-124z_2^3+8z_2^4)\right\}x^4-4z\left\{1-z(5+18z_2)-z^2(8+35z_2-120z_2^2)\right.\nonumber\\
&&\hspace{-1em}\left.+8z^3(6-17z_2+65z_2^2-44z_2^3)-48z^4z_2(9-8z_2)(1-z_2)^2\right\}x^3+2z^2\left\{3-2z(11+17z_2)\right. \nonumber\\
&&\hspace{-1em}\left. +32z^2(1-2z_2)^2+32z^3(1-9z_2+15z_2^2-5z_2^3)\right\}x^2-4z^3\left\{1-z(8+5z_2)\right. \nonumber\\
&&\hspace{-1em} \left.\left.+8z^2(2-11z_2-z_2^2)+48z^3z_2(1-z_2)^2\right\}x +z^4\left\{1-8z(1+z_2)+16z^2(1-4z_2+z_2^2)\right\}\right]~. \nonumber
\end{eqnarray}
\noindent$\underline{\mbox{{\bf Example 3:}}~K_{{\IF}_2}}$\\
We use the mirror curve (\ref{eqn:4.48}). The one-holed torus amplitude is
\begin{eqnarray}
\label{eqn:D.5}%ラベル指定
W^{(1,1)}(x)&=&-\frac{32z^4dx}{\sqrt{\sigma(x)}\Delta(z_{\alpha})}\left(2G(\tau)^2+f_1^{(1,1)}(x)G(\tau)+f_0^{(1,1)}(x)\right)~, \\
f_1^{(1,1)}(x)&=&-\frac{z^4x(x^2-x+z_0^{-\frac12}z)}{\sigma(x)}\Delta_0(z_{\alpha})~,\quad \Delta_0(z_{\alpha})=(1-4z_0^{-\frac12}z)^2-64z^2~, \nonumber\\
f_0^{(1,1)}(x)&=&\frac{z^4}{9\sigma(x)^2}\bigl[\bigl\{(1-4z_0^{-\frac12}z)^2-96z^2\bigl\}x^8-4\bigl\{(1-4z_0^{-\frac12}z)^2-96z^2\bigl\}x^7+2\bigl\{3-22z_0^{-\frac12}z \nonumber\\
&&\hspace{-1em} +32z_0^{-1}z^2(1-9z_0)+32z_0^{-\frac32}z^3(1-6z_0)\bigl\}x^6-4\bigl\{1-5z_0^{-\frac12}z-8z_0^{-1}z^2(1+9z_0) \nonumber\\
&&\hspace{-1em} +48z_0^{-\frac32}z^3(1-10z_0)+384z_0^{-1}z^4(1-4z_0)\bigl\}x^5+\bigl\{1+4z_0^{-\frac12}-2z_0^{-1}z^2(37-44z_0) \nonumber\\
&&\hspace{-1em} +16z_0^{-\frac32}z^3(9-164z_0)+32z_0^{-2}z^4(3+74z_0-360z_0^2)\bigl\}x^4-4z_0^{-\frac12}z \bigl\{1-z_0^{-\frac12}z(5-29z_0) \nonumber\\
&&\hspace{-1em} -4z_0^{-1}z^2(2+79z_0)+16z_0^{-\frac32}z^3(3+5z_0-108z_0^2)+192z_0^{-1}z^4(1-4z_0)\bigl\}x^3+2z_0^{-1}z^2\bigl\{3+14z_0 \nonumber\\
&&\hspace{-1em} -2z_0^{-\frac12}z(11+48z_0)+32z_0^{-1}z^2(1-6z_0-24z_0^2)+32z_0^{-\frac32}z^3(1-4z_0)(1+6z_0)\bigl\}x^2 \nonumber\\
&&\hspace{-1em} -4z_0^{-\frac32}z^3\bigl\{1+5z_0-4z_0^{-\frac12}z(2+13z_0-12z_0^2)+16z_0^{-1}z^2(1-4z_0)(1+9z_0) \nonumber\\
&&\hspace{-1em} -192z_0^{-\frac32}z^3(1-4z_0)^2\bigl\}x+z_0^{-2}z^4(1-4z_0)\bigl\{(1-4z_0^{-\frac12}z)^2-96z^2\bigl\}\bigl]~. \nonumber
\end{eqnarray}
In appendix E, we will compute the A-model amplitues $F^{(1,1)}$ and $F^{(0,3)}$ on $K_{{\IF}_2}$ and ${\IC}^3/{\IZ}_4$.

\section{Topological open string amplitudes on $K_{{\relax{\rm I\kern-.15em F}}_2}$ and ${\IC}^3/{\IZ}_4$}%secE

In this appendix, we summarize the results of the topological open string amplitudes on $K_{{\IF}_2}$ and ${\IC}^3/{\IZ}_4$. Let us collect the necessary data for these computation. We use the mirror curve (\ref{eqn:4.48}),
\begin{equation}
y^2-(x^2-x+{\overline z}_0^{-\frac12}z)y+z^2=0~,\quad \sigma(x)=(x^2-x+{\overline z}_0^{-\frac12}z)^2-4z^2=\prod_{i=1}^4(x-s_i)~,
\label{eqn:E.1}%ラベル指定
\end{equation}
and from (\ref{eqn:4.51}) the closed mirror maps are given by
\begin{equation}
{\overline T}_0=\log {\overline z}_0-2\log \frac{1+\sqrt{1-4{\overline z}_0}}{2}~,\quad T_z=\log z+2\sum_{m,n \ge 0,~m \ge 2n}^{\infty}\frac{(2m-1)!}{m!n!^2(m-2n)!}z^m{\overline z}_0^{-\frac{m}{2}+n}~,
\label{eqn:E.2}%ラベル指定
\end{equation}
where we must reparametrize the variables as (\ref{eqn:4.30}) and (\ref{eqn:4.31}),
\begin{eqnarray}
\label{eqn:E.3}%ラベル指定
&&{\overline T}_0=T_0~,\quad T_z=T_5+\frac12 T_0~,\\
&&{\overline z}_0=z_0~,\quad z=z_0^{\frac12}z_5~.
\label{eqn:E.4}%ラベル指定
\end{eqnarray}
Furthermore we consider the framing ambiguity by (\ref{eqn:C.5}) and (\ref{eqn:C.7}),
\begin{equation}
x=x_f^{-1}y_f^f~,\quad y_f=1-x_f-(f-{\overline z}_0^{-\frac12}z)x_f^2-\frac{f}{2}(3f-6{\overline z}_0^{-\frac12}z+1)x_f^3+\cdots~,
\label{eqn:E.5}%ラベル指定
\end{equation}
and the open mirror map is given by the same form as (\ref{eqn:4.38}),
\begin{equation}
X_f=-\left(\frac{z}{Q_z}\right)^{\frac12}x_f~.
\label{eqn:E.6}%ラベル指定
\end{equation}
Let us compute the open string amplitudes on $K_{{\IF}_2}$. The disk amplitude $F^{(0,1)}$ is obtained from the Abel-Jacobi map as follows \cite{AgVa} ;\vspace{-0.05em}
\begin{eqnarray}
F^{(0,1)}&=&\int^{x_f} \log y_f \frac{dx_f}{x_f} \nonumber\\
&=& \Bigl\{1+Q_5(1+Q_0)+3Q_0Q_5^2+5Q_0Q_5^3+7Q_0Q_5^4+5Q_0^2Q_5^3+9Q_0Q_5^5+35Q_0^2Q_5^4+\cdots \Bigr\}X_f \nonumber\\
&&\hspace{-1em} -\Bigl\{\frac14(2f+1)+fQ_5(1+Q_0)+\frac14(2f+1)Q_5^2+2(2f+1)Q_0Q_5^2+\frac14(2f+1)Q_0^2Q_5^2+\cdots \Bigr\}X_f^2 \nonumber\\
&&\hspace{-1em} +\Bigl\{\frac{1}{18}(3f+1)(3f+2)+\frac{f}{2} (3f+1)Q_5(1+Q_5+Q_0)+\frac12(3f+1)(5f+2)Q_0Q_5^2+\cdots \Bigr\}X_f^3 \nonumber\\
&&\hspace{-1em} -\Bigl\{\frac{1}{48}(2f+1)(4f+1)(4f+3)+\frac{f}{3}(2f+1)(4f+1)Q_5(1+Q_0)+\frac{f}{4}(4f+1)^2Q_5^2+\cdots \Bigr\}X_f^4+\cdots~. \nonumber
\end{eqnarray}
The annulus amplitude $F^{(0,2)}$ is obtained from (\ref{eqn:2.13}) and (\ref{eqn:4.50}) as follows ;\vspace{-0.05em}
\begin{eqnarray}
F^{(0,2)}&=&\int^{x_{f,1}}\int^{x_{f,2}} B \left(x_1(x_{f,1}),x_2(x_{f,2})\right)-\frac{dx_{f,1}dx_{f,2}}{(x_{f,1}-x_{f,2})^2} \nonumber\\
&=& \Bigl\{\frac{f}{2}(f+1)+f^2Q_5(1+Q_0)+\frac{f}{2}(f+1)Q_5^2(1+Q_0^2)+2(2f^2+2f+1)Q_0Q_5^2(1+2Q_5+2Q_0Q_5) \nonumber\\
&&\hspace{2em}+6(2f^2+2f+1)Q_0Q_5^4(1+\frac43 Q_5+Q_0^2)+\frac92(11f^2+11f+8)Q_0^2Q_5^4+\cdots\Bigr\}X_{f,1}X_{f,2} \nonumber\\
&&\hspace{-4em} -(2f+1)\Bigl\{\frac{f}{3}(f+1)+f^2Q_5(1+Q_0+Q_5)+(5f^2+3f+1)Q_0Q_5^2+\frac{f}{3}(f+1)Q_5^3+f^2Q_0^2Q_5^2 \nonumber\\
&&\hspace{-1em}+3(4f^2+3f+1)Q_0Q_5^3(1+Q_0)+\frac{f}{3}(f+1)Q_0^3Q_5^3+5(4f^2+3f+1)Q_0Q_5^4(1+Q_0^2)+\cdots \Bigr\}X_{f,1}^2X_{f,2} \nonumber\\
&&\hspace{-4em} +(3f+1)\Bigl\{\frac{f}{8}(f+1)(3f+2)+\frac{f^2}{2}(3f+2)Q_5(1+Q_0)+\frac{3f^2}{4}(3f+1)Q_5^2+(9f^3+9f^2+4f+1)Q_0Q_5^2 \nonumber\\
&&\hspace{-1em}+\frac{f^2}{4}Q_5^2\left(3(3f+1)Q_0^2+2(3f+2)Q_5\right)+\frac12(51f^3+58f^2+24f+4)Q_0Q_5^3(1+Q_0)+\cdots \Bigr\}X_{f,1}^3X_{f,2} \nonumber\\
&&\hspace{-4em} +(2f+1)\Bigl\{\frac{f}{4}(f+1)(2f+1)+f^2(2f+1)Q_5(1+Q_0)+\frac{f^2}{2}(6f+1)Q_5^2+(2f+1)(6f^2+2f+1)Q_0Q_5^2 \nonumber\\
&&\hspace{-1em}+\frac{f^2}{2}Q_5^2\left((6f+1)Q_0^2+2(2f+1)Q_5\right)+(2f+1)(17f^2+8f+2)Q_0Q_5^3(1+Q_0)+\cdots \Bigr\}X_{f,1}^2X_{f,2}^2 +\cdots~. \nonumber
\end{eqnarray}
These results completely agree with the topological vertex calculus in appendix A by
\begin{equation}
f=-{\widetilde f}-1~,
\label{eqn:E.7}%ラベル指定
\end{equation}
as far as we have checked. Next we compute the higher amplitudes $F^{(1,1)}$ from (\ref{eqn:D.5}) and $F^{(0,3)}$ from (\ref{eqn:4.62}). Because the higher amplitudes depend on the positions of the ramification points $q_i$ of the mirror curve as seen from (\ref{eqn:2.3}), we cannot compute the framed amplitudes by the reparametrization (\ref{eqn:4.5}) of the open string moduli for (\ref{eqn:D.5}) and (\ref{eqn:4.62}). So here we only compute the zero-framing amplitudes as follows\footnote{Note that we use the reparametrized mirror curve (\ref{eqn:E.1}) as $x \to x^{-1},~y \to x^{-2}y$ which does not change the positions of the ramification points, and $|\omega(q)-\omega({\bar q})|$ is invariant under this reparametrization, so we can compute these amplitudes by this simple reparametrization.} ;
\begin{eqnarray}
F^{(1,1)}&=&\int^x W^{(1,1)}(x^{-1}) \frac{-1}{x^2}\nonumber\\
&=& -\frac{1}{24}\left(1+Q_5+Q_0Q_5+3Q_0Q_5^2+5Q_0Q_5^3+7Q_0Q_5^4+5Q_0^2Q_5^3+9Q_0Q_5^5-157Q_0^2Q_5^4+\cdots \right)X \nonumber\\
&&\hspace{-1em} +\frac{1}{24}\left(1+Q_5^2-4Q_0Q_5^2-8Q_0Q_5^3+Q_0^2Q_5^2-12Q_0Q_5^4-8Q_0^2Q_5^3-16Q_0Q_5^5-213Q_0^2Q_5^4+\cdots \right)X^2 \nonumber\\
&&\hspace{-1em} -\frac{1}{24}\left(1-7Q_0Q_5^2+Q_5^3-21Q_0Q_5^3-35Q_0Q_5^4-21Q_0^2Q_5^3-49Q_0Q_5^5-284Q_0^2Q_5^4+Q_0^3Q_5^3+\cdots \right)X^3 \nonumber\\
&&\hspace{-1em} +\frac{1}{24}\left(1-14Q_0Q_5^2-28Q_0Q_5^3+Q_5^4-56Q_0Q_5^4-28Q_0^2Q_5^3-84Q_0Q_5^5-348Q_0^2Q_5^4+\cdots \right)X^4 +\cdots~, \nonumber\\
F^{(0,3)}&=&\int^{x_1}\int^{x_2}\int^{x_3} W^{(0,3)}(x_1^{-1},x_2^{-1},x_3^{-1}) \frac{-1}{x_1^2x_2^2x_3^2}\nonumber\\
&=& \left(Q_0Q_5^2+3Q_0Q_5^3+5Q_0Q_5^4+3Q_0^2Q_5^3+7Q_0Q_5^5+35Q_0^2Q_5^4+5Q_0^3Q_5^4+182Q_0^3Q_5^5+\cdots \right)X_1X_2X_3 \nonumber\\
&&\hspace{-2.5em} -\left(Q_0Q_5^2+2Q_0Q_5^3+4Q_0Q_5^4+2Q_0^2Q_5^3+6Q_0Q_5^5+24Q_0^2Q_5^4+126Q_0^2Q_5^5+4Q_0^3Q_5^4+\cdots \right)X_1^2X_2X_3 \nonumber\\
&&\hspace{-2.5em} +\left(Q_0Q_5^2+2Q_0Q_5^3+3Q_0Q_5^4+2Q_0^2Q_5^3+5Q_0Q_5^5+21Q_0^2Q_5^4+105Q_0^2Q_5^5+3Q_0^3Q_5^4+\cdots \right)X_1^3X_2X_3 \nonumber\\
&&\hspace{-2.5em} +\left(Q_0Q_5^2+2Q_0Q_5^3+3Q_0Q_5^4+2Q_0^2Q_5^3+5Q_0Q_5^5+18Q_0^2Q_5^4+90Q_0^2Q_5^5+3Q_0^3Q_5^4+\cdots \right)X_1^2X_2^2X_3 \nonumber\\
&&\hspace{-2.5em} -\left(Q_0Q_5^2+2Q_0Q_5^3+3Q_0Q_5^4+2Q_0^2Q_5^3+4Q_0Q_5^5+17Q_0^2Q_5^4+72Q_0^2Q_5^5+3Q_0^3Q_5^4+\cdots \right)X_1^2X_2^2X_3^2+\cdots~. \nonumber
\end{eqnarray}
These results completely agree with the topological vertex calculus with the framing ${\widetilde f}=-1$ in appendix A as far as we have checked. Here let us move to the orbifold phase ${\IC}^3/{\IZ}_4$, and we compute the open string amplitudes in this phase. This phase transition is carried out by a transfomation ${\overline z}_0=a_3^{-2},~z=a_1^{-2}$, and the orbifold mirror maps are given by (\ref{eqn:4.53}) and (\ref{eqn:4.55}). As above, in this orbifold phase we obtain the disk amplitude $F_o^{(0,1)}$ and the annulus amplitude $F_o^{(0,2)}$ with the framing $f \in {\IZ}$, and the one-holed torus amplitude $F_o^{(1,1)}$ and the genus zero, three-hole amplitude $F_o^{(0,3)}$ with the zero-framing.\footnote{The disk amplitude $F_o^{(0,1)}$ with the framing $f=0$ completely agrees with the result in \cite{BrTan}(version 4). The other amplitudes $F_o^{(0,2)}$, $F_o^{(1,1)}$ and $F_o^{(0,3)}$ also completely agree with the computation via the BKMP's remodeling approach by A. Brini \cite{BrCav} (see footnote 4 in subsection 4.4).} Note that as pointed out in \cite{BKMP2} it is not necessary that $f$ is integer, rather, from a viewpoint of the A-model it is natural that $f$ has a fractional value in the orbifold phase.
\begin{eqnarray}
F_o^{(0,1)}&=& \Bigl\{\Bigl(1-\frac{1}{16}\frac{s_{1/2}^2}{2!}-\frac{3}{256}\frac{s_{1/2}^4}{4!}+\cdots \Bigr)s_{1/4}+\Bigl(\frac{1}{32}s_{1/2}+\frac{9}{512}\frac{s_{1/2}^3}{3!}+\frac{321}{8192}\frac{s_{1/2}^5}{5!}+\cdots \Bigr)\frac{s_{1/4}^3}{3!} \nonumber\\
&&\hspace{1em} -\Bigl(\frac{3}{64}+\frac{37}{1024}\frac{s_{1/2}^2}{2!}+\frac{1551}{16384}\frac{s_{1/2}^4}{4!}+\cdots \Bigr)\frac{s_{1/4}^5}{5!}+\cdots \Bigr\}X_f \nonumber\\
&&\hspace{-1em} +\Bigl\{\Bigl(\frac{1}{2}s_{1/2}-\frac{1}{8}\frac{s_{1/2}^3}{3!}+\frac{1}{32}\frac{s_{1/2}^5}{5!}+\cdots \Bigr)+(2f+1)\Bigl(-\frac12+\frac{1}{16}\frac{s_{1/2}^2}{2!}+\frac{7}{512}\frac{s_{1/2}^6}{6!}+\cdots \Bigr)\frac{s_{1/4}^2}{2!} \nonumber\\
&&\hspace{1em} -(2f+1)\Bigl(\frac{1}{16}s_{1/2}+\frac{3}{128}\frac{s_{1/2}^3}{3!}+\frac{27}{512}\frac{s_{1/2}^5}{5!}+\cdots \Bigr)\frac{s_{1/4}^4}{4!}+\cdots \Bigr\}X_f^2 \nonumber\\
&&\hspace{-1em} +(3f+1)\Bigl\{\Bigl(-\frac{1}{3}s_{1/2}+\frac{7}{48}\frac{s_{1/2}^3}{3!}-\frac{41}{768}\frac{s_{1/2}^5}{5!}+\cdots \Bigr)s_{1/4}+\Bigl(\frac{1}{3}(3f+2)-\frac{1}{48}(9f+7)\frac{s_{1/2}^2}{2!}+\cdots \Bigr)\frac{s_{1/4}^3}{3!} \nonumber\\
&&\hspace{1em} +\Bigl(\frac{1}{192}(60f+43)s_{1/2}+\frac{1}{1024}(60f+73)\frac{s_{1/2}^3}{3!}+\frac{1}{49152}(8460f+11963)\frac{s_{1/2}^5}{5!}+\cdots \Bigr)\frac{s_{1/4}^5}{5!}+\cdots \Bigr\}X_f^3 \nonumber\\
&&\hspace{-1em} +\Bigl\{-\frac14 \Bigl(1+(4f+1)\frac{s_{1/2}^2}{2!}-(4f+1)\frac{s_{1/2}^4}{4!}+\cdots\Bigr)+(2f+1)(4f+1)\Bigl(\frac12 s_{1/2}-\frac{5}{16}\frac{s_{1/2}^3}{3!}+\cdots\Bigr)\frac{s_{1/4}^2}{2!} \nonumber\\
&&\hspace{1em} +(2f+1)(4f+1)\Bigl(-\frac12(4f+3)+\frac12(f+1)\frac{s_{1/2}^2}{2!}-\frac{1}{64}(12f+7)\frac{s_{1/2}^4}{4!}+\cdots\Bigr)\frac{s_{1/4}^4}{4!}+\cdots\Bigr\}X_f^4+\cdots~, \nonumber\\
F_o^{(0,2)}&=& \Bigl\{\Bigl(-\frac14(4f+1)s_{1/2}+\frac{1}{32}(8f+1)\frac{s_{1/2}^3}{3!}+\cdots \Bigr)+\Bigl(\frac{1}{8}(8f^2+8f+1)-\frac{f}{8}(f+1)\frac{s_{1/2}^2}{2!} +\cdots \Bigr)\frac{s_{1/4}^2}{2!} \nonumber\\
&&\hspace{1em} +\Bigl(\frac{1}{128}(16f^2+16f-1)s_{1/2}+\frac{3}{128}(2f^2+2f-1)\frac{s_{1/2}^3}{3!}+\cdots \Bigr)\frac{s_{1/4}^4}{4!} +\cdots \Bigr\}X_{f,1}X_{f,2} \nonumber\\
&&\hspace{-1em} +(2f+1)\Bigl\{\Bigl(\frac{1}{8}(8f+1)s_{1/2}-\frac{1}{128}(56f+5)\frac{s_{1/2}^3}{3!}+\cdots \Bigr)s_{1/4}+\Bigl(-\frac{1}{16}(32f^2+32f+3)+\cdots \Bigr)\frac{s_{1/4}^3}{3!} \nonumber\\
&&\hspace{1em} -\Bigl(\frac{1}{512}(320f^2+344f+3)s_{1/2}+\frac{1}{8192}(960f^2+1752f-255)\frac{s_{1/2}^3}{3!}+\cdots \Bigr)\frac{s_{1/4}^5}{5!}+\cdots \Bigr\}X_{f,1}^2X_{f,2} \nonumber\\
&&\hspace{-1em} +(3f+1)\Bigl\{\Bigl(\frac13+\frac{1}{6}(6f+1)\frac{s_{1/2}^2}{2!}+\cdots \Bigr)+\Bigl(-\frac{1}{24}(72f^2+54f+5)s_{1/2}+\cdots \Bigr)\frac{s_{1/4}^2}{2!} \nonumber\\
&&\hspace{1em} +\Bigl(\frac{1}{4}(36f^3+60f^2+27f+2)-\frac{1}{192}(432f^3+864f^2+414f+19)\frac{s_{1/2}^2}{2!}+\cdots\Bigr)\frac{s_{1/4}^4}{4!} +\cdots \Bigr\}X_{f,1}^3X_{f,2} \nonumber\\
&&\hspace{-1em} +(2f+1)\Bigl\{\Bigl(\frac12+f \frac{s_{1/2}^2}{2!}-f \frac{s_{1/2}^4}{4!}+\cdots \Bigr)+(2f+1)\Bigl(-\frac{1}{8}(16f+1)s_{1/2}+\frac{1}{16}(20f+1)\frac{s_{1/2}^3}{3!}+\cdots \Bigr)\frac{s_{1/4}^2}{2!} \nonumber\\
&&\hspace{1em} +(2f+1)\Bigl(\frac{3}{8}(16f^2+16f+1)-\frac{1}{64}(96f^2+128f+5)\frac{s_{1/2}^2}{2!}+\cdots \Bigr)\frac{s_{1/4}^4}{4!}+\cdots \Bigr\}X_{f,1}^2X_{f,2}^2+\cdots~, \nonumber\\
F_o^{(1,1)}&=& \Bigl\{-\Bigl(\frac{1}{48}+\frac{1}{768}\frac{s_{1/2}^2}{2!}+\frac{65}{12288}\frac{s_{1/2}^4}{4!}+\cdots\Bigr)s_{1/4}+\Bigl(\frac{5}{1536}s_{1/2}+\frac{81}{8192}\frac{s_{1/2}^3}{3!}+\frac{19145}{393216}\frac{s_{1/2}^5}{5!}+\cdots \Bigr)\frac{s_{1/4}^3}{3!} \nonumber\\
&&\hspace{1em} -\Bigl(\frac{9}{1024}+\frac{1367}{49152}\frac{s_{1/2}^2}{2!}+\frac{40345}{262144}\frac{s_{1/2}^4}{4!}+\cdots \Bigr)\frac{s_{1/4}^5}{5!}+\cdots \Bigr\}X \nonumber\\
&&\hspace{-1em} +\Bigl\{\Bigl(-\frac{1}{48}s_{1/2}+\frac{5}{384}\frac{s_{1/2}^3}{3!}-\frac{1}{768}\frac{s_{1/2}^5}{5!}+\cdots\Bigr)+\Bigl(\frac{1}{32}-\frac{1}{192}\frac{s_{1/2}^2}{2!}+\frac{5}{3072}\frac{s_{1/2}^4}{4!}+\frac{11}{2048}\frac{s_{1/2}^6}{6!}+\cdots\Bigr)\frac{s_{1/4}^2}{2!} \nonumber\\
&&\hspace{1em} +\Bigl(\frac{1}{512}s_{1/2}-\frac{5}{512}\frac{s_{1/2}^3}{3!}-\frac{1277}{24576}\frac{s_{1/2}^5}{5!}+\cdots\Bigr)\frac{s_{1/4}^4}{4!}+\cdots\Bigr\}X^2 \nonumber\\
&&\hspace{-1em} +\Bigl\{\Bigl(\frac{5}{144}s_{1/2}-\frac{53}{2304}\frac{s_{1/2}^3}{3!}+\frac{235}{12288}\frac{s_{1/2}^5}{5!}+\cdots\Bigr)s_{1/4}+\Bigl(-\frac{1}{12}+\frac{59}{2304}\frac{s_{1/2}^2}{2!}-\frac{305}{18432}\frac{s_{1/2}^4}{4!}+\cdots\Bigr)\frac{s_{1/4}^3}{3!} \nonumber\\
&&\hspace{1em} +\Bigl(-\frac{155}{3072}s_{1/2}+\frac{1903}{49152}\frac{s_{1/2}^3}{3!}+\frac{716555}{2359296}\frac{s_{1/2}^5}{5!}+\cdots\Bigr)\frac{s_{1/4}^5}{5!}+\cdots\Bigr\}X^3 \nonumber\\
&&\hspace{-1em} +\Bigl\{\Bigl(-\frac13+\frac{1}{24}\frac{s_{1/2}^2}{2!}-\frac{7}{96}\frac{s_{1/2}^4}{4!}+\frac{31}{384}\frac{s_{1/2}^6}{6!}+\cdots\Bigr)+\Bigl(-\frac{3}{32}s_{1/2}+\frac{11}{128}\frac{s_{1/2}^3}{3!}-\frac{233}{3072}\frac{s_{1/2}^5}{5!}+\cdots\Bigr)\frac{s_{1/4}^2}{2!} \nonumber\\
&&\hspace{1em} +\Bigl(\frac{5}{16}-\frac{39}{256}\frac{s_{1/2}^2}{2!}+\frac{3}{32}\frac{s_{1/2}^4}{4!}+\frac{403}{2048}\frac{s_{1/2}^6}{6!}+\cdots\Bigr)\frac{s_{1/4}^4}{4!}+\cdots\Bigr\}X^4+\cdots~,\nonumber\\
F_o^{(0,3)}&=& \Bigl\{\Bigl(\frac{1}{32}s_{1/2}-\frac{7}{512}\frac{s_{1/2}^3}{3!}-\frac{79}{8192}\frac{s_{1/2}^5}{5!}+\cdots\Bigr)s_{1/4}+\Bigl(-\frac{3}{64}+\frac{11}{1024}\frac{s_{1/2}^2}{2!}+\frac{353}{16384}\frac{s_{1/2}^4}{4!}+\cdots\Bigr)\frac{s_{1/4}^3}{3!} \nonumber\\
&&\hspace{1em} -\Bigl(\frac{33}{2048}s_{1/2}+\frac{1989}{32768}\frac{s_{1/2}^3}{3!}+\frac{218993}{524288}\frac{s_{1/2}^5}{5!}+\cdots\Bigr)\frac{s_{1/4}^5}{5!}+\cdots\Bigr\}X_1X_2X_3 \nonumber\\
&&\hspace{-1em} +\Bigl\{\Bigl(\frac{1}{2}+\frac{1}{16}\frac{s_{1/2}^2}{2!}-\frac{1}{32}\frac{s_{1/2}^4}{4!}+\cdots\Bigr)+\Bigl(-\frac{1}{16}s_{1/2}+\frac{3}{128}\frac{s_{1/2}^3}{3!}-\frac{7}{512}\frac{s_{1/2}^5}{5!}+\cdots\Bigr)\frac{s_{1/4}^2}{2!} \nonumber\\
&&\hspace{1em} +\Bigl(\frac{9}{64}-\frac{7}{256}\frac{s_{1/2}^2}{2!}-\frac{3}{512}\frac{s_{1/2}^4}{4!}+\cdots\Bigr)\frac{s_{1/4}^4}{4!}+\cdots\Bigr\}X_1^2X_2X_3 \nonumber\\
&&\hspace{-1em} +\Bigl\{\Bigl(-\frac{2}{3}-\frac{1}{16}\frac{s_{1/2}^2}{2!}+\frac{7}{64}\frac{s_{1/2}^4}{4!}+\cdots\Bigr)s_{1/4}+\Bigl(\frac{35}{192}s_{1/2}-\frac{127}{1024}\frac{s_{1/2}^3}{3!}+\frac{385}{16384}\frac{s_{1/2}^5}{5!}+\cdots\Bigr)\frac{s_{1/4}^3}{3!} \nonumber\\
&&\hspace{1em} +\Bigl(-\frac{19}{32}+\frac{599}{3072}\frac{s_{1/2}^2}{2!}+\frac{71}{1024}\frac{s_{1/2}^4}{4!}+\cdots\Bigr)\frac{s_{1/4}^5}{5!}+\cdots\Bigr\}X_1^3X_2X_3 \nonumber\\
&&\hspace{-1em} +\Bigl\{\Bigl(-\frac34-\frac{1}{64}\frac{s_{1/2}^2}{2!}+\frac{65}{1024}\frac{s_{1/2}^4}{4!}+\cdots\Bigr)s_{1/4}+\Bigl(\frac{15}{128}s_{1/2}-\frac{177}{2048}\frac{s_{1/2}^3}{3!}+\frac{895}{32768}\frac{s_{1/2}^5}{5!}+\cdots\Bigr)\frac{s_{1/4}^3}{3!} \nonumber\\
&&\hspace{1em} +\Bigl(-\frac{111}{256}+\frac{603}{4096}\frac{s_{1/2}^2}{2!}+\frac{1245}{65536}\frac{s_{1/2}^4}{4!}+\cdots\Bigr)\frac{s_{1/4}^5}{5!}+\cdots\Bigr\}X_1^2X_2^2X_3 \nonumber\\
&&\hspace{-1em} +\Bigl\{\Bigl(-\frac12 s_{1/2}+\frac18 \frac{s_{1/2}^2}{2!}-\frac{1}{32}\frac{s_{1/2}^5}{5!}+\cdots\Bigr)+\Bigl(\frac{7}{4}-\frac{1}{8}\frac{s_{1/2}^2}{2!}-\frac{15}{128}\frac{s_{1/2}^4}{4!}+\cdots\Bigr)\frac{s_{1/4}^2}{2!}+\cdots\Bigr\}X_1^2X_2^2X_3^2+\cdots~. \nonumber
\end{eqnarray}
These orbifold amplitudes are the ${\IZ}_2$ and ${\IZ}_4$ monodromy invariant discussed in subsection 4.4, and these results predict the open orbifold Gromov-Witten invariants of ${\IC}^3/{{\IZ}_4}$.


\begin{thebibliography}{99}
 \addcontentsline{toc}{section}{References}
\bibitem{CHSW} P. Candelas, G. Horowitz, A. Strominger and E. Witten,
\textquotedblleft Vacuum configurations for superstrings,\textquotedblright~
Nucl. Phys. B {\bf 258} (1985) 46-74.
%
\bibitem{BCOV1} M. Bershadsky, S. Cecotti, H. Ooguri and C. Vafa,
\textquotedblleft Kodaira-Spencer theory of gravity and exact results for quantum string amplitudes,\textquotedblright~
Commun. Math. Phys. {\bf 165} (1994) 311-428 [arXiv:hep-th/9309140].
%
\bibitem{AGNT1} I. Antoniadis, E. Gava, K. S. Narain and T. R. Taylor,
\textquotedblleft Topological amplitudes in string theory,\textquotedblright~
Nucl. Phys. B {\bf 413} (1994) 162-184 [arXiv:hep-th/9307158].
%
\bibitem{AGNT2} I. Antoniadis, E. Gava, K. S. Narain and T. R. Taylor,
\textquotedblleft ${\cal N}=2$ type II-Heterotic duality and higher derivative F-terms,\textquotedblright~
Nucl. Phys. B {\bf 455} (1995) 109-130 [arXiv:hep-th/9507115].
%
\bibitem{GoVa1} R. Gopakumar and C. Vafa,
\textquotedblleft M-theory and topological strings-I,\textquotedblright~
arXiv:hep-th/9809187.
%
\bibitem{GoVa2} R. Gopakumar and C. Vafa,
\textquotedblleft M-theory and topological strings-II,\textquotedblright~
arXiv:hep-th/9812127.
%
\bibitem{KKVa} S. Katz, A. Klemm and C. Vafa,
\textquotedblleft Geometric engineering of quantum field theories,\textquotedblright~
Nucl. Phys. B {\bf 497} (1997) 173-195 [arXiv:hep-th/9609239].
%
\bibitem{IqKP1} A. Iqbal and A.-K. Kashani-Poor,
\textquotedblleft Instanton counting and Chern-Simons theory,\textquotedblright~
Adv. Theor. Math. Phys. {\bf 7} (2004) 457-497 [arXiv:hep-th/0212279].
%
\bibitem{IqKP2} A. Iqbal and A.-K. Kashani-Poor,
\textquotedblleft $SU(N)$ geometries and topological string amplitudes,\textquotedblright~
Adv. Theor. Math. Phys. {\bf 10} (2006) 1-32 [arXiv:hep-th/0306032].
%
\bibitem{EgKn1} T. Eguchi and H. Kanno,
\textquotedblleft Topological strings and Nekrasov's formulas,\textquotedblright~
JHEP {\bf 0312} (2003) 006 [arXiv:hep-th/0310235].
%
\bibitem{EgKn2} T. Eguchi and H. Kanno,
\textquotedblleft Geometric transitions, Chern-Simons gauge theory and Veneziano type amplitudes,\textquotedblright~
Phys. Lett. B {\bf 585} (2004) 163-172 [arXiv:hep-th/0312234].
%
\bibitem{Nek} N. Nekrasov,
\textquotedblleft Seiberg-Witten prepotential from instanton counting,\textquotedblright~
Adv. Theor. Math. Phys. {\bf 7} (2004) 831-864 [arXiv:hep-th/0206161].
%
\bibitem{Coat} T. Coates,
\textquotedblleft Wall-crossings in toric Gromov-Witten theory II: local examples,\textquotedblright~
arXiv:0804.2592 [math.AG].
%
\bibitem{AgKMVa} M. Aganagic, A. Klemm, M. Mari\~no and C. Vafa,
\textquotedblleft The topological vertex,\textquotedblright~
Commun. Math. Phys. {\bf 254} (2005) 425-478 [arXiv:hep-th/0305132].
%
\bibitem{COGP} P. Candelas, X. de la Ossa, P. Green and L. Parkes,
\textquotedblleft A pair of Calabi-Yau manifolds as an exactly soluble superconformal theory,\textquotedblright~
Nucl. Phys. B {\bf 359} (1991) 21-74.
%
\bibitem{Wal2} J. Walcher,
\textquotedblleft Extended holomorphic anomaly and loop amplitudes in open topological string,\textquotedblright~
Nucl. Phys. B {\bf 817} (2009) 167-207 [arXiv:0705.4098 [hep-th]].
%
\bibitem{Mar1} M. Mari\~no,
\textquotedblleft Open string amplitudes and large order behavior in topological string theory,\textquotedblright~
JHEP {\bf 0803} (2008) 060 [arXiv:hep-th/0612127].
%
\bibitem{BKMP1} V. Bouchard, A. Klemm, M. Mari\~no and S. Pasquetti,
\textquotedblleft Remodeling the B-model,\textquotedblright~
Commun. Math. Phys. {\bf 287} (2009) 117-178 [arXiv:0709.1453 [hep-th]].
%
\bibitem{EO1} B. Eynard and N. Orantin,
\textquotedblleft Invariants of algebraic curves and topological expansion,\textquotedblright~
Commun. Number Theor. Phys. {\bf 1} (2007) 347-452 [arXiv:math-ph/0702045].
%
\bibitem{EMO} B. Eynard, M. Mari\~no and N. Orantin,
\textquotedblleft Holomorphic anomaly and matrix models,\textquotedblright~
JHEP {\bf 0706} (2007) 058 [arXiv:hep-th/0702110].
%
\bibitem{DV2} R. Dijkgraaf and C. Vafa,
\textquotedblleft Two dimensional Kodaira-Spencer theory and three dimensional Chern-Simons gravity,\textquotedblright~
arXiv:0711.1932 [hep-th].
%
\bibitem{BKMP2} V. Bouchard, A. Klemm, M. Mari\~no and S. Pasquetti,
\textquotedblleft Topological open strings on orbifolds,\textquotedblright~
arXiv:0807.0597 [hep-th].
%
\bibitem{BCOV2} M. Bershadsky, S. Cecotti, H. Ooguri and C. Vafa,
\textquotedblleft Holomorphic anomalies in topological field theories,\textquotedblright~
Nucl. Phys. B {\bf 405} (1993) 279-304 [arXiv:hep-th/9302103].
%
\bibitem{KlZas} A. Klemm and E. Zaslow,
\textquotedblleft Local mirror symmetry at higher genus,\textquotedblright~
arXiv:hep-th/9906046.
%
\bibitem{Bat} V. V. Batyrev,
\textquotedblleft Dual polyhedra and mirror symmetry for Calabi-Yau hypersurfaces in toric varieties,\textquotedblright~
J. Alg. Geom. {\bf 3} (1994) 493-535 [arXiv:alg-geom/9310003].
%
\bibitem{Mina} S. Minabe,
\textquotedblleft Topological vertex and its applications,\textquotedblright~
Ph.D. Thesis at the Graduate School of Mathematics, Nagoya University [http://hdl.handle.net/2237/7811].
%
\bibitem{CKYZ} T.-M. Chiang, A. Klemm, S.-T. Yau and E. Zaslow,
\textquotedblleft Local mirror symmetry: calculations and interpretations,\textquotedblright~
Adv. Theor. Math. Phys. {\bf 3} (1999) 495-565 [arXiv:hep-th/9903053].
%
\bibitem{ACKM} J. Ambj{\o}rn, L. Chekhov, C. F. Kristjansen and Yu. Makeenko,
\textquotedblleft Matrix model calculations beyond the spherical limit,\textquotedblright~
Nucl. Phys. B {\bf 404} (1993) 127-172; Erratum-ibid. B {\bf 449} (1995) 681 [arXiv:hep-th/9302014].
%
\bibitem{Ake} G. Akemann,
\textquotedblleft Higher genus correlators for the hermitian matrix model with multiple cuts,\textquotedblright~
Nucl. Phys. B {\bf 482} (1996) 403-430 [arXiv:hep-th/9606004].
%
\bibitem{Eyn1} B. Eynard,
\textquotedblleft Topological expansion for the 1-hermitian matrix model correlation functions,\textquotedblright~
JHEP {\bf 0411} (2004) 031 [arXiv:hep-th/0407261].
%
\bibitem{Mar2} M. Mari\~no,
\textquotedblleft Chern-Simons theory and topological strings,\textquotedblright~
Rev. Mod. Phys. {\bf 77} (2005) 675-720 [arXiv:hep-th/0406005].
%
\bibitem{AgVa} M. Aganagic and C. Vafa,
\textquotedblleft Mirror symmetry, D-branes and counting holomorphic disks,\textquotedblright~
arXiv:hep-th/0012041.
%
\bibitem{CoKa} D. Cox and S. Katz,
\textquotedblleft Mirror symmetry and algebraic geometry,\textquotedblright~
American Mathematical Society, 1999.
%
\bibitem{HoVa} K. Hori and C. Vafa,
\textquotedblleft Mirror symmetry,\textquotedblright~
arXiv:hep-th/0002222.
%
\bibitem{FoJi} B. Forbes and M. Jinzenji,
\textquotedblleft Extending the Picard-Fuchs system of local mirror symmetry,\textquotedblright~
J. Math. Phys. {\bf 46} (2005) 082302 [arXiv:hep-th/0503098].
%
\bibitem{DV1} R. Dijkgraaf and C. Vafa,
\textquotedblleft Matrix models, topological strings, and supersymmetric gauge theories,\textquotedblright~
Nucl. Phys. B {\bf 644} (2002) 3-20 [arXiv:hep-th/0206255].
%
\bibitem{Mar3} M. Mari\~no,
\textquotedblleft Les Houches lectures on matrix models and topological strings,\textquotedblright~
arXiv:hep-th/0410165.
%
\bibitem{LerM} W. Lerche and P. Mayr,
\textquotedblleft On ${\cal N}=1$ mirror symmetry for open type II strings,\textquotedblright~
arXiv:hep-th/0111113.
%
\bibitem{Fay} J. D. Fay,
\textquotedblleft Theta functions on Riemann surfaces,\textquotedblright~
Lecture Note in Mathematics {\bf 352}, Springer, 1973.
%
\bibitem{BrTan} A. Brini and A. Tanzini,
\textquotedblleft Exact results for topological strings on resolved $Y(p,q)$ singularities,\textquotedblright~
Commun. Math. Phys. {\bf 289} (2009) 205-252 [arXiv:0804.2598 [hep-th]].
%
\bibitem{HKR} B. Haghighat, A. Klemm and M. Rauch,
\textquotedblleft Integrability of the holomorphic anomaly equations,\textquotedblright~
JHEP {\bf 0810} (2008) 097 [arXiv:0809.1674 [hep-th]].
%
\bibitem{ALM} M. Alim, J.D. L\"ange and P. Mayr,
\textquotedblleft Global properties of topological string amplitudes and orbifold invariants,\textquotedblright~
arXiv:0809.4253 [hep-th].
%
\bibitem{ABK} M. Aganagic, V. Bouchard and A. Klemm,
\textquotedblleft Topological strings and (almost) modular forms,\textquotedblright~
Commun. Math. Phys. {\bf 277} (2008) 771-819 [arXiv:hep-th/0607100].
%
\bibitem{AKMV} M. Aganagic, A. Klemm, M. Mari\~no and C. Vafa,
\textquotedblleft Matrix model as a mirror of Chern-Simons theory,\textquotedblright~
JHEP {\bf 0402} (2004) 010 [arXiv:hep-th/0211098].
%
\bibitem{BrCav} A. Brini and R. Cavalieri, to appear.
%
\bibitem{HuKl0} M. x. Huang and A. Klemm,
\textquotedblleft  Holomorphic anomaly in gauge theories and matrix models,\textquotedblright~
JHEP {\bf 0709} (2007) 054 [arXiv:hep-th/0605195].
%
\bibitem{HuKl} M. x. Huang and A. Klemm,
\textquotedblleft Holomorphicity and modularity in Seiberg-Witten theories with matter,\textquotedblright~
arXiv:0902.1325 [hep-th].
%
\bibitem{Chek} L. Chekhov,
\textquotedblleft Genus one correlation to multi-cut matrix model solutions,\textquotedblright~
Theor. Math. Phys. {\bf 141} (2004) 1640-1653; Teor. Mat. Fiz. {\bf 141} (2004) 358-374 [arXiv:hep-th/0401089].
%
\bibitem{GhV} D. Ghoshal and C. Vafa,
\textquotedblleft $c=1$ string as the topological theory of the conifold,\textquotedblright~
Nucl. Phys. B {\bf 453} (1995) 121-128 [arXiv:hep-th/9506122].
%
\bibitem{HKQ} M. x. Huang, A. Klemm and S. Quackenbush,
\textquotedblleft Topological string theory on compact Calabi-Yau: modularity and boundary conditions,\textquotedblright~
Lect. Notes Phys. {\bf 757} (2009) 45-102 [arXiv:hep-th/0612125].
%
\bibitem{KMT} A. Klemm, M. Mari\~no and S. Theisen,
\textquotedblleft Gravitational corrections in supersymmetric gauge theory and matrix models,\textquotedblright~
JHEP {\bf 0303} (2003) 051 [arXiv:hep-th/0211216].
%
\end{thebibliography}
\end{document}